\documentclass[preprint,noshowpacs,noshowkeys,floatfix,aps,pra]{revtex4-1}
\usepackage{multirow}
\usepackage{graphicx}

\begin{document}

\title{Conditional quasi-exact solvability of the quantum planar pendulum and of its anti-isospectral hyperbolic counterpart}

\author{Simon Becker}
\author{Marjan Mirahmadi}
\author{Burkhard Schmidt}
\email{burkhard.schmidt@fu-berlin.de}
\affiliation{
Institute for Mathematics, Freie Universit\"{a}t Berlin \\ Arnimallee 6, D-14195 Berlin, Germany}

\author{Konrad Schatz}
\author{Bretislav Friedrich}
\email{bretislav.friedrich@fhi-berlin.mpg.de}
\affiliation{
Fritz-Haber-Institut der Max-Planck-Gesellschaft \\ Faradayweg 4-6, D-14195 Berlin, Germany}

\date{\today}

\begin{abstract}

We have subjected the  planar pendulum eigenproblem to a symmetry analysis with the goal of explaining the relationship between its conditional quasi-exact solvability (C-QES) and the topology of its eigenenergy surfaces,  established in our earlier work [Frontiers in Physical Chemistry and Chemical Physics {\bf 2}, 1-16, (2014)]. The present analysis revealed that this relationship can be traced to the structure of the tridiagonal matrices representing the symmetry-adapted pendular Hamiltonian, as well as enabled us to identify many more -- forty in total to be exact -- analytic solutions. Furthermore,  an analogous analysis of the hyperbolic counterpart of the planar pendulum, the Razavy problem, which was shown to be also C-QES [American Journal of Physics {\bf 48}, 285 (1980)], confirmed that it is anti-isospectral with the pendular eigenproblem. Of key importance for both eigenproblems proved to be the topological index $\kappa$, as it determines the loci of the intersections (genuine and avoided) of the eigenenergy surfaces spanned by the dimensionless interaction parameters $\eta$ and $\zeta$. It also encapsulates the conditions under which analytic solutions to the two eigenproblems obtain and provides the number of analytic solutions. At a given $\kappa$, the anti-isospectrality occurs for single states only (i.e., not for doublets), like C-QES  holds solely for integer values of $\kappa$, and only occurs for the lowest eigenvalues of the pendular and Razavy Hamiltonians, with the order of the eigenvalues reversed for the latter.  For all other states, the pendular and Razavy spectra become in fact qualitatively different, as higher pendular states appear as doublets whereas all higher Razavy states are singlets. 

\end{abstract}

\maketitle

\section{Introduction}
\label{intro}

Like the harmonic oscillator, the planar pendulum is key to the understanding of a number of prototypical one-dimensional problems in chemistry and physics, partly listed in Table I. 
However, unlike the harmonic oscillator problem, the planar pendulum one is not analytically
(or exactly) solvable, i.e., its Schr\"{o}dinger equation does {\it not} possess  algebraic
solutions that cover the entire spectrum of the problem's Hamiltonian. Instead, the problem
is only conditionally quasi-exactly solvable  \cite{dutra1993,lopez1993}, i.e., its algebraic solutions only exist for finitely many eigenvalues of the pendular Hamiltonian (quasi-exact solvability, QES), and, moreover, only obtain if the problem's interaction parameters
satisfy a particular set of conditions (conditional quasi-exact solvability, C-QES).
Previous work \cite{Schmidt2014b} has identified {\it some} analytic solutions and conditions for a planar pendulum whose potential is comprised of a trigonometric expansion up to second order, sometimes referred to as the square planar pendulum \cite{Child_2007}. 
Below, by planar pendulum we always mean the square planar pendulum.\footnote{We note that the analytic asymptotic states that the planar pendulum possesses are exempt from our considerations here as these are eigenstates of a different Hamiltonian.} 

Herein we seek to extend the batch of the analytic solutions of the planar pendulum problem by making use of the connection, recognized in our previous work \cite{Schmidt2014b}, between the topology of the eigenenergy surfaces and the conditional quasi-solvability, as well as of the symmetry of the problem and the properties of its {\it anti-isospectral} \cite{Krajewska1997a, Khare1998a} counterpart. 
Thereby we identify a range of analytic wavefunctions endowed with a clear physical meaning and pertaining to both periodic and aperiodic single- as well as multiple-well potentials.   

We start by invoking the analytic solutions of the planar pendulum problem found earlier via supersymmetric quantum mechanics (SUSY QM \cite{Cooper1995f}) and reported in Ref. \cite{Schmidt2014b}. 
There it is shown how transformations between pairs of (almost) isospectral Hamiltonians can be used to construct analytic solutions for Schr\"{o}dinger equations which are otherwise hard to find.
In our present work these solutions are classified into four categories, each of them associated with one of the four irreducible representations of the $C_{2v}$ point group. 
For each of the irreducible representations, the Hamiltonian of the planar pendulum is found to be an infinite tridiagonal matrix containing a finite-dimensional block characterised by a particular condition imposed on the pendulum's parameters and expressed in terms of an integer, termed the {\it topological index}. 
The value of the topological index is related to the dimension of the finite block and provides the number of analytic solutions. 
In principle, there are arbitrarily many values of the topological index and hence infinitely many analytic solutions within a given irreducible representation.  
Apart from the trigonometric potential of the planar pendulum, we also investigate its hyperbolic counterpart,  known as the Razavy potential \cite{Razavy1980m}, which obtains via an anti-isospectral transformation of the pendular potential. 
The Razavy potential\footnote{Also known as the double sinh-Gordon (DSGH) potential.} is related to the symmetric double Morse potential. 
Its applications are listed in Table I. 

Like in the pendular case, the Razavy Hamiltonian becomes tridiagonal in the irreducible representations of its symmetry group. 
However, its symmetry is that of the  $C_{i}$ point group, yielding just two irreducible representations. 
As shown below, the intersections of the trigonometric (pendular) and hyperbolic (Razavy) spectra as functions of the interaction parameters yield analytic  eigenenergies corresponding to the analytic solutions. 
This is in agreement with the properties of the energy levels of the spin system
formulations of both the planar pendulum and the Razavy Hamiltonians 
\cite{Turbiner1988a,Ulyanov1984v,Leon2014a,Konwent1998h, Finkel1999f}. 
In either case, we obtain the conditions for quasi-analytic solvability (QES) as a trivial consequence of our approach, independent of previous algebraic work, see e.g., Refs.~
\cite{Turbiner1988a,Turbiner1999a,Finkel1999f,Gomez2007d,Gomez2005d,Djakov2005p}. 

Finally, we take advantage of the spectral properties of the Schr\"{o}dinger equation of the planar pendulum, which corresponds to a periodic Sturm-Liouville differential equation known as the Whittaker-Hill equation \cite{Magnus2004w,Roncaratti2010l,Hemery2010a,Finkel1999f,Leon2014a,Djakov2005p}, as well as of the properties of its anti-isospectral transform to gain an insight into the eigenproperties of both the planar pendulum and Razavy systems. What we found is that outside the range of C-QES, the higher states are all doublets (pendulum) or singlets (Razavy system).

This paper is organised as follows: 
In Section~\ref{sec:ham}, we review the general properties of the planar pendulum as well as the Razavy Hamiltonians. 
In Section~\ref{sec:solve}, the conditions for quasi-analytic solvability are studied for either of the two potentials, with a particular attention to their symmetry; at the same time, we investigate the analytic solutions of the Schr\"{o}dinger equation for both Hamiltonians and their mutual relationship. 
A brief survey of the numerical solutions of the Schr\"{o}dinger equation for the two systems is given in Section~\ref{sec:num}. 
Finally, Section~\ref{sec:con} provides a summary of the present work.

\section{Properties of the Hamiltonians}
\label{sec:ham}

In this Section we describe the properties of the planar pendulum and Razavy
Hamiltonians whose respective potentials are related via an anti-isospectral transformation.

\subsection{Planar pendulum}
\label{sec:ham_t}

We consider the Hamiltonian of the planar quantum pendulum to be of the form
\begin{equation}
\label{eq:ham_t}
H_t= - \frac{d^2}{d\theta^2} + V_{t}(\theta)
\end{equation}
where all energies are expressed in units of the rotational constant $B\equiv \hbar^2/(2I)$ with $I$ being the moment of inertia.
The periodic potential   
\begin{equation}
\label{eq:pot_t}
V_t(\theta)= -\eta \cos\theta-\zeta \cos^{2}\theta
\end{equation}
is a {\it trigonometric} series (hence the subscript {\it t}) up to second order for angle  $\theta \in (0, 2\pi)$ whose Fourier terms are weighted by the (real) dimensionless parameters $\eta$ and $\zeta$. 
For $\eta = \zeta = 0$, Hamiltonian (\ref{eq:ham_t}) becomes that of a free rotor or a particle on a ring.
Throughout this work we consider $\zeta>0$; we note that the structure of the solutions is qualitatively different for negative values of $\zeta$ \cite{Roncaratti2010l}.
For a discussion of positive and negative values of $\eta$, see below. 

\begin{figure}
\includegraphics[width=8cm]{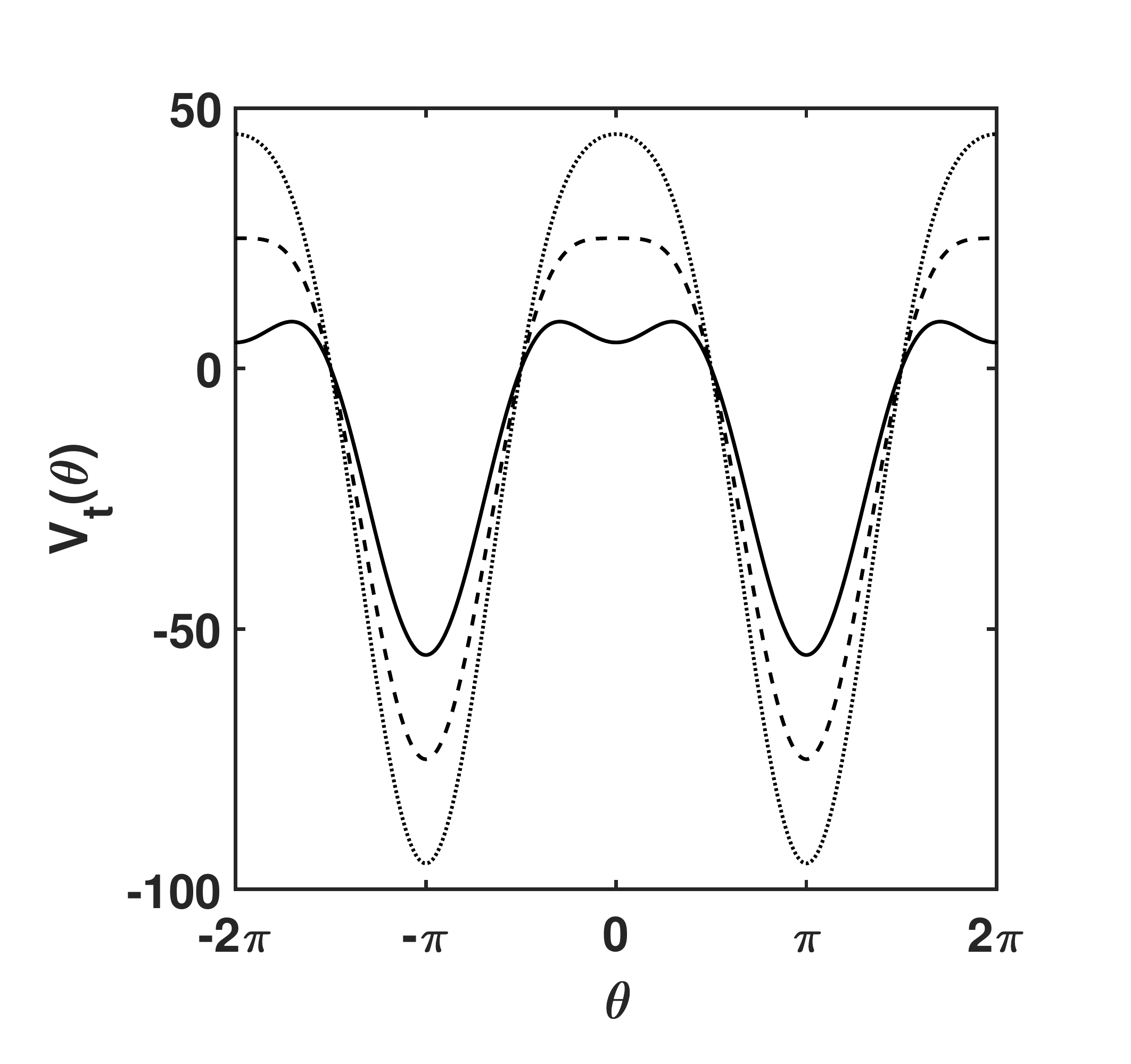}
\caption{Planar pendulum (trigonometric) potential, Eq. (\ref{eq:pot_t}), for $\zeta=25$ and $\eta=-30$ (full curve), $\eta=-50$ (dashed curve), and $\eta=-70$ (dotted curve). Note that the potential is a double well for $\left|\eta\right| < \left|2\zeta\right|$ and a single well for $\left|\eta\right| \geq \left|2\zeta\right|$. }
\label{fig:pendulum}
\end{figure}

The Schr\"{o}dinger equation
\begin{equation}
\label{eq:se_t}
-\frac{d^2\psi_{t}(\theta)}{d\theta^2}- \left[\eta \cos\theta + \zeta \cos^{2}\theta\right] \psi_{t}(\theta) = E_{t} \psi_{t}(\theta)
\end{equation}
reduces for either $\eta=0$ or $\zeta=0$ to a Mathieu equation \cite{Friedrich_1991,Leibscher-Schmidt_2009,Schmidt2014b}. We note that Mathieu equations do not have analytic solutions but possess many analytic properties \cite{Stegun}.
The pendular  potential  (\ref{eq:pot_t}) is  $2\pi$-periodic and  for $\theta \in (0, 2\pi)$ assumes a shape 
that depends in the following way on the relative magnitude of $|\eta|$ and $2\zeta$:
\begin{itemize}
	\item For $\left|\eta\right| < 2\zeta$ and $\eta<0$, $V_t$ consists of an asymmetric double well with a global minimum of $ (\eta-\zeta)$ at $\theta_{min,g}=\pi$, a local minimum of $(-\eta - \zeta)$ at $\theta_{min,l}=0$, and global maxima of $\frac{\eta^2}{4\zeta}$ at $\theta_{max}= \arccos\left[-\frac{\eta}{2\zeta}\right]$, $2\pi-\arccos\left[-\frac{\eta}{2\zeta}\right]$, see Figure \ref{fig:pendulum}. 
For $\eta>0$, the potential consists of an asymmetric double well with a local minimum of $(\eta - \zeta)$ at $\theta_{min,l}=\pi$, global minima of $-(\eta + \zeta)$ at $\theta_{min,g}=0, 2\pi$, and global maxima of $\frac{\eta^2}{4\zeta}$ at $\theta_{max}= \arccos\left[-\frac{\eta}{2\zeta}\right]$, $2\pi-\arccos\left[-\frac{\eta}{2\zeta}\right]$. 

\item For $\left|\eta\right| > 2\zeta$ and $\eta<0$, $V_t$ is a single well with a minimum of  $ (\eta-\zeta) $ at $\theta_{min}=\pi$ and a maximum of $(-\eta - \zeta)$ at $\theta_{max}=0,2\pi$, see Figure \ref{fig:pendulum}. For $\eta>0$, $V_t$ is a single well with a minimum of  $ -(\eta+\zeta) $ at $\theta_{min}=0,2\pi$ and a maximum of $(\eta - \zeta)$ at $\theta_{max}=\pi$. We note that for $\left|\eta\right|= 2\zeta$, the maxima become flat, as a result of which
the first three derivatives vanish at $\theta_{max}$.
\end{itemize}

As can be gleaned from Figure \ref{fig:pendulum}, potential (\ref{eq:pot_t}) is invariant under the transformations $\theta \mapsto \theta + 2\pi$ and $\theta \mapsto -\theta$. As a consequence, the planar pendulum possesses  a  symmetry isomorphic with that of the point group $C_{2v}$ (with $\theta \mapsto \theta + 2\pi$ and $\theta \mapsto -\theta$ corresponding, respectively, to rotation and inversion). Below we exploit this symmetry by making use of its irreducible representations to simplify the Hamiltonian matrix. Apart from considering $2\pi$-periodic wavefunctions on the $\theta \in (0, 2\pi)$ interval, we also consider $4\pi$-periodic wavefunctions on the $\theta \in (-2\pi, 2\pi)$ interval that are $2\pi$-antiperiodic and thus are not solutions of the pendular eigenproblem, Eq. (\ref{eq:se_t}). We include these wavefunctions as they may prove useful for tackling problems involving Berry's geometric phase \cite{Berry_1988}.\footnote {We note that a mapping $\theta \mapsto \frac{\theta}{2}$  would make the $A$ ($B$) solutions periodic (anti-periodic) in $\pi$, see below. This could be of interest in treatments of, e.g. (circular) motion of an atom around a hetero-nuclear diatomic. }

\subsection{Razavy system}
\label{sec:ham_h}
 
The quasi-exactly solvable Schr\"{o}dinger equation for a symmetric double-well potential introduced by Razavy \cite{Razavy1980m,Razavy2003m} can be recast in the form
\begin{equation}
\label{eq:se_h}
	-\frac{d^2 \psi_{h}(x)}{d x^2}+ \left(\eta \cosh x + \zeta
	\cosh^{2}x\right)\psi_{h}(x)=  E_{h} \psi_{h}(x) 
\end{equation}
where $x$ is a linear coordinate, $x \in {(-\infty,\infty})$. The eigenvalues $E_h$ and eigenfunctions $\psi_h$ of Eq. (\ref{eq:se_h}) are labeled with the subscript $h$ to indicate that they pertain to Razavy's {\it hyperbolic} potential,
\begin{equation} 
\label{eq:pot_h}
	V_h(x) =\eta \cosh x + \zeta \cosh^{2}x \,. 
\end{equation}
We note that the eigenproblems for the planar pendulum, Eq. (\ref{eq:ham_t}), and the Razavy system, Eq.~(\ref{eq:se_h}), are related by the anti-isospectral transformation (AIS) that maps
\begin{eqnarray}
\label{eq:ais}
	 \theta & \mapsto & \mathrm{i} x \nonumber \\
	E_t & \mapsto{}& - E_h \,.
\end{eqnarray}
However, the planar pendulum and the Razavy systems are anti-isospectral only over a finite range of their spectra $E_t$ and $E_h$, as will be described in detail below.

The Razavy  potential  (\ref{eq:pot_h}) exhibits minima only for $\zeta>0$. Their general shape depends on the parameters $\eta$ and $\zeta$ in the following way:
\begin{itemize}

\item For $\eta<0$ and $\left|\eta\right| >2 \zeta$, $V_h$ is a symmetric double well whose minima of $-\frac{\eta^2}{4\zeta}$ occur at $x = \pm \mathrm{arccosh}(-\frac{\eta}{2\zeta})$
 and its local maximum of ($\eta + \zeta$) at $x=0$. For $\eta<0$ and $\left|\eta\right| \le 2 \zeta$, $V_h$ a single well potential with a minimum of $(\eta+\zeta)$ at $x=0$.
\item For $\eta>0$, $V_h$ is a single well (irrespective of the relative magnitude of $\eta$ and $\zeta$) with a minimum of $(\eta + \zeta)$ at $x=0$. If, in addition, $\left|\eta\right| = 2\zeta$, the well has a flat bottom with the first three derivatives vanishing at the minimum.		
\end{itemize}
 
For $|\eta| \gg \zeta$, the Razavy potential approaches the shape of a double-Morse potential with a flat barrier \cite{Konwent1998h}. 
Using $\mathrm{arccosh}(y) = \ln (y + \sqrt{y^2 - 1} )$, the separation of the Morse wells is given by $2\ln(-\eta/\zeta)$. 

We note that the Razavy potential (\ref{eq:pot_h}) is only invariant under the parity transformation $x \mapsto -x$ (as well as under the transformation $x\mapsto x + 2i\pi$) and thus has the symmetry of the point group $C_i$,  which is a subgroup of $C_{2v}$. This fact will help us to elucidate the connections between the planar pendulum and Razavy systems.

In order to bring into play the Razavy potential as a double-well potential, we need to consider $\eta<0$ (and $\zeta>0$, as before). Under such conditions, however,  whenever the Razavy potential is a (symmetric) double-well potential, namely for $\left|\eta\right| > 2\zeta$, the pendular potential is a single-well potential. And conversely, under the same conditions, whenever the Razavy potential is a single-well potential, namely for $\left|\eta\right| < 2\zeta$, the pendular potential is an (asymmetric) double well potential. 

\section{Conditional quasi-exact solvability}
\label{sec:solve}
In this section we investigate the symmetries of the solution
spaces of the planar pendulum and Razavy systems and relate them to 
the conditions of quasi-analytic solvability.

\subsection{Symmetries and seed functions}
\label{sec:solve_symm}
 
We map the symmetry operations of the $C_{2v}$ point group \cite{Bunker2005} onto those of the planar pendulum (trigonometric) system in the following way:

\begin{eqnarray}
\label{eq:map_C2v}
E & \mapsto & E\equiv R(4\pi)  \nonumber \\
C_2 & \mapsto & R(2\pi) \nonumber \\
\sigma_v(xz) & \mapsto & P(\vartheta=0) \nonumber \\
\sigma_v(yz) & \mapsto & P(\vartheta=\pi)
\end{eqnarray}
with $E$ the identity operation, $R(\vartheta)$ rotation by angle $\vartheta$ and $P(\vartheta)$ the parity operation, $\theta-\vartheta \mapsto -\theta-\vartheta$, with  $\vartheta$ the origin; $\sigma$ stands for reflection from a plane.
As we are interested in both $2\pi$-periodic and anti-periodic wavefunctions, the angle $\theta$ is considered to be in the ($-2\pi,2\pi$) domain. 

\begin{figure}
\includegraphics[width=17cm]{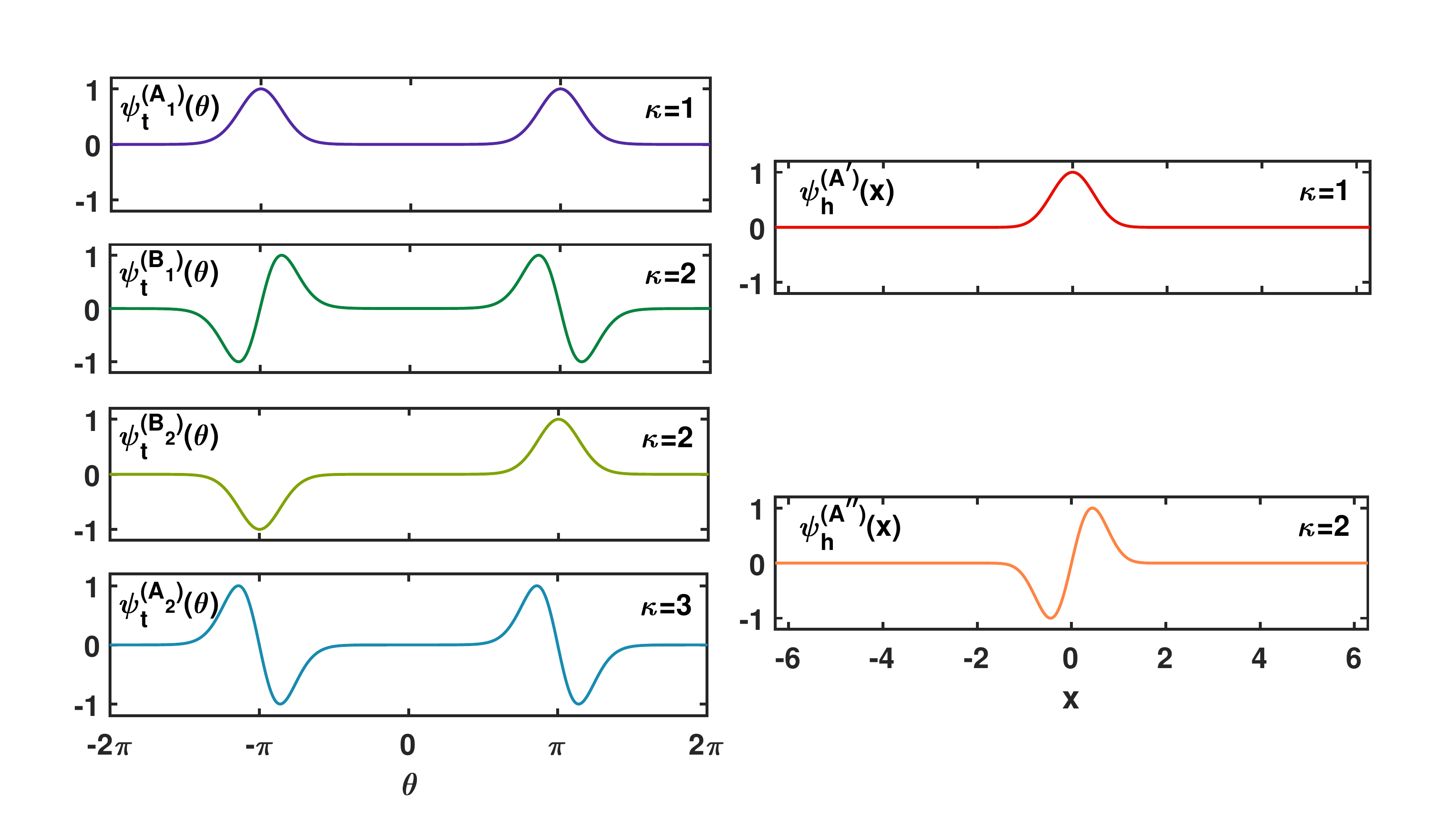} 
\caption{Seed wavefunctions listed in Table \ref{tab:seed}, with $\beta=-5$ for both the planar pendulum trigonometric system (left) and the Razavy hyperbolic system (right). Note that the color coding introduced here is used throughout the paper.}
\label{fig:seed}
\end{figure}

For the Razavy (hyperbolic) system, the mapping of the  $C_i$ point group is
\begin{eqnarray}
\label{eq:map_Ci}
E & \mapsto & E  \nonumber \\
i & \mapsto  & P
\end{eqnarray}
where $E$ is the identity and $P$ the parity operation, $x \mapsto -x$. Table \ref{tab:irreps} provides a summary of the characters of the irreducible representations $\Gamma_t$ and $\Gamma_h$ for both the planar pendulum and Razavy systems. 
Indeed, the analytic solutions found so far, see Refs. \cite{Schmidt2014b} and  \cite{Razavy1980m}, for the lowest states of the two systems exhibit, respectively, the presumed $C_{2v}$ and  $C_i$ symmetries. 
The eigenenergies and wavefunctions of these states are listed in Table \ref{tab:seed} along with their symmetry labels $\Gamma_t$ or $\Gamma_h$. 
The corresponding wavefunctions are also shown in Figure \ref{fig:seed}, whose inspection allows to verify at once the assignment of the symmetry labels. 

The $C_i$ point group is a subgroup of $C_{2v}$, whose irreducible representations $A_1$, $B_1$  and $A_2$, $B_2$ correlate, respectively, with the irreducible representation $A'$  and $A''$ of $C_i$. The parity operation $P$, Eq. (\ref{eq:map_C2v}), applied to the hyperbolic system plays the role of the $P(0)$ operation, Eq. (\ref{eq:map_Ci}), applied to the trigonometric system. 

As an aside, we note that the totally symmetric trigonometric wavefunction, $\psi_{t,1}^{(A_1)}\propto \exp(\beta\cos\theta)$, see Table \ref{tab:seed} and  Figure \ref{fig:seed}, has the form of the von Mises distribution \cite{Mises_1918}, which is the circular analog of a normal distribution (or a Gaussian wavepacket). 
Although the latter is omnipresent in quantum mechanics textbooks, the former is hardly mentioned in the literature at all as a solution of Schr\"{o}dinger's equation (\ref{eq:se_t}). The same can be said about the hyperbolic analog of the von Mises distribution, $\psi_{h,1}^{(A')}\propto \exp(\beta\cosh x)$, which is a solution of Eq.~(\ref{eq:se_h}). 
The lack of attention to these as well as all the other analytic solutions listed in Table \ref{tab:seed} and shown in Figure \ref{fig:seed} may be due to the fact that these solutions only obtain for certain integer values of 
\begin{equation}
\label{eq:kappa}
\kappa \equiv \frac{|\eta|}{\sqrt{\zeta}} \equiv \frac{|\eta|}{|\beta|}
\end{equation}
where $\beta$ is a short-hand for $\pm \sqrt{\zeta}$. Hence $\eta=\kappa \beta$ and $\zeta=\beta^2$. 
A given value of $\kappa$ defines a particular condition for the quasi-exact solvability of either the planar pendulum or Razavy problems which, therefore, belong to the class of conditionally quasi-exactly solvable systems.
As expanded upon below, the (integer) value of the index $\kappa$ also serves to specify the number of analytic solutions obtainable. For more details, see Section~\ref{sec:solve}.
At the same time, as described in Ref. \cite{SchmiFri_2014,Schmidt2014b}, the index $\kappa$ characterises the structure/topology of the pendulum's eigenenergy surfaces, which is why it was termed in Ref. \cite{SchmiFri_2014} the {\it topological index}. 

Table \ref{tab:seed} also reveals that the analytic eigenvalues of the planar pendulum and Razavy problems exhibit anti-isospectrality as well as a correspondence between the eigenfunctions pertaining to a given eigenvalue and its counterpart upon replacing $\cos \mapsto \cosh$ (or $\sin \mapsto \sinh$). Below, we show that these correspondences remain in place for all analytic solutions for the two potentials in question. This is a manifestation of the ``duality property'', which entails that quasi-exactly solvable problems arise in pairs of different forms whose analytic eigenenergies coincide, up to a change of sign \cite{Krajewska1997a}.

Below we make use of the analytic wavefunctions listed in Table \ref{tab:seed} and shown in Figure \ref{fig:seed} as seed functions that allow us to find, in principle, arbitrarily many additional analytic solutions. 

\subsection{Planar pendulum}
\label{sec:solve_t} 

By making use of Eq.~(\ref{eq:kappa}) and the substitution 
\begin{equation}
\label{eq:subst_t1}
\psi_t(\theta) = f_t(\theta)\exp(\beta\cos\theta)
\end{equation}
the original Schr\"{o}dinger equation (\ref{eq:se_t}) for the planar pendulum becomes
\begin{equation}
\label{eq:ince_t}
-\frac{d^2f_{t}(\theta)}{d\theta^2}+2 \beta\sin\theta \frac{df_{t}(\theta)}{d\theta} - \left[\beta^2 + \beta (\kappa-1)\cos\theta\right] f_{t}(\theta)=E_{t} f_{t}(\theta)
\end{equation}
which is the equation of Ince \cite{Olver_2010}. Each of its four nontrivial periodic solutions \cite{Magnus2004w} corresponds  to one of the symmetries of the planar pendulum: even and $2\pi$-periodic solution corresponds to the $A_1$ symmetry; odd and $2\pi$-periodic solution to $A_2$, even and $2\pi$-antiperiodic solution to $B_1$, and odd and $2\pi$ anti-periodic solution to $B_2$, see also \cite{Hemery2010a, Djakov2005p, Roncaratti2010l}. 

With the further substitution
\begin{equation}
\label{eq:subst_t2}
u \equiv  \cos\frac{\theta}{2}
\end{equation}
the Ince Eq.~(\ref{eq:ince_t}) can be written as 
\begin{eqnarray}
\label{eq:T_t}
T_{t,\kappa} \phi_{t,\kappa}&\equiv&\frac{1}{4}(1-u^{2})\frac{d^2\phi_{t,\kappa}}{du^2} + \left(2\beta u -2\beta u^3-\frac{u}{4}\right)\frac{d\phi_{t,\kappa}}{du}+ \left((2u^2-1)(\kappa-1)\beta+\beta^2\right)\phi_{t,\kappa} \nonumber \\
&=& -E_{t,\kappa}\phi_{t,\kappa}
\end{eqnarray}
where $T_{t,\kappa}$ is the negative of the Schr\"{odinger} operator of the planar pendulum  that depends parametrically on the topological index $\kappa$ and where $\phi_{t,\kappa}(u)$ is equivalent to $f_t(\theta)$ for a given value of $\kappa$.
The last substitution has served to eliminate all trigonometric functions; as a result,  from here on we only have to deal with polynomials in the new argument $u$.
The two transformations (\ref{eq:subst_t1}) and (\ref{eq:subst_t2}) can now also be applied to the four trigonometric seed functions given in the left part of Table~\ref{tab:seed} (and shown in Fig.~\ref{fig:seed}), yielding the following expressions
\begin{eqnarray}
\label{eq:seed_t}
\phi_{t,1}^{(A_1)}(u) &=& 1 \nonumber \\
\phi_{t,2}^{(B_1)}(u) &=& u \nonumber \\
\phi_{t,2}^{(B_2)}(u) &=& \pm \sqrt{1-u^2} \nonumber \\
\phi_{t,3}^{(A_2)}(u) &=& u \sqrt{1-u^2} \,.
\end{eqnarray}
These lowest-order eigenfunctions that transform according to the irreducible representations of the $C_{2v}$ point group can be used to symmetry-adapt the Schr\"{odinger} operator $T_{t,\kappa}$, Eq. (\ref{eq:T_t}), to the symmetry of the planar pendulum via the following gauge transformation,
\begin{equation}
\label{eq:gauge_t}
T_{t,\kappa}^{(\Gamma_t)} \equiv \frac{1}{\phi_{t,\kappa}^{(\Gamma_t)}}T_{t,\kappa}\phi_{t,\kappa}^{(\Gamma_t)}
\end{equation}
with $\Gamma_t\in \left\{A_1,B_1,B_2,A_2\right\}$ and where $\kappa\in\{1,2,2,3\}$ as given in Eq.~\ref{eq:seed_t}. 
Note that the structure of the Lie algebras (from which the symmetry-adapted operators could have been constructed as well) is left invariant by this gauge transformation, as is the spectrum \cite{kamran1990,Glopez1994d}.

In order to obtain explicit matrix representations of the $T_{t,\kappa}^{(\Gamma_t)}$ operators, we make use of a basis set of monomials in $u$
\begin{equation}
\label{eq:basis_t}
\left\{1,u^2,u^4,\ldots\right\}
\end{equation}
comprised of even-order powers only. 
These are totally symmetric (pertaining to the $A_1$ irreducible representation) with respect to the symmetry operations of the planar pendulum as given by Eq.~(\ref{eq:map_C2v}) and listed in Tab.~\ref{tab:irreps} and thus not affecting the symmetry of the $T^{(\Gamma_t)}_{t,\kappa}$ operators. 

In the basis set (\ref{eq:basis_t}), the four symmetry-adapted Schr\"{o}dinger operators of Eq.~(\ref{eq:gauge_t}) are represented by tridiagonal matrices with the following superdiagonal matrix elements
\begin{eqnarray}
\langle u^{2\ell-2} | T_{t,\kappa}^{(A_1)} | u^{2\ell} \rangle &=& \ell^{2}-\ell/2 \nonumber \\
\langle u^{2\ell-2} | T_{t,\kappa}^{(B_1)} | u^{2\ell} \rangle &=& \ell^{2}+\ell/2 \nonumber \\
\langle u^{2\ell-2} | T_{t,\kappa}^{(B_2)} | u^{2\ell} \rangle &=& \ell^{2}-\ell/2 \nonumber \\
\langle u^{2\ell-2} | T_{t,\kappa}^{(A_2)} | u^{2\ell} \rangle &=& \ell^{2}+\ell/2 \label{eq:T_sup}
\end{eqnarray}
for natural numbers $\ell$. The superdiagonals are always non-negative for $\ell>0$. 
The diagonal elements are given by
\begin{eqnarray}
\langle u^{2\ell} | T_{t,\kappa}^{(A_1)} | u^{2\ell} \rangle &=& \beta^{2} - \ell^2              + 4\beta \ell -(\kappa-1)\beta \nonumber \\
\langle u^{2\ell} | T_{t,\kappa}^{(B_1)} | u^{2\ell} \rangle &=& \beta^{2} - \ell^2 - \ell - 1/4 + 4\beta \ell -(\kappa-3)\beta \nonumber \\
\langle u^{2\ell} | T_{t,\kappa}^{(B_2)} | u^{2\ell} \rangle &=& \beta^{2} - \ell^2 - \ell - 1/4 + 4\beta \ell -(\kappa-1)\beta \nonumber \\
\langle u^{2\ell} | T_{t,\kappa}^{(A_2)} | u^{2\ell} \rangle &=& \beta^{2} - \ell^2 -2\ell -  1  + 4\beta \ell -(\kappa-3)\beta \label{eq:T_diag}
\end{eqnarray}
with integer $\ell \ge 0$.
The subdiagonal elements are
\begin{eqnarray}
\langle u^{2\ell} | T_{t,\kappa}^{(A_1)} | u^{2\ell-2} \rangle &=& 4\beta \left(-\ell + \frac{\kappa+1}{2} \right)  \nonumber \\
\langle u^{2\ell} | T_{t,\kappa}^{(B_1)} | u^{2\ell-2} \rangle &=& 4\beta \left(-\ell + \frac{\kappa  }{2} \right)  \nonumber \\
\langle u^{2\ell} | T_{t,\kappa}^{(B_2)} | u^{2\ell-2} \rangle &=& 4\beta \left(-\ell + \frac{\kappa  }{2} \right)  \nonumber \\
\langle u^{2\ell} | T_{t,\kappa}^{(A_2)} | u^{2\ell-2} \rangle &=& 4\beta \left(-\ell + \frac{\kappa-1}{2} \right)  \label{eq:T_sub}
\end{eqnarray} 
with integer $\ell>0$. 

Thus, by virtue of the substitutions (\ref{eq:subst_t1}) and (\ref{eq:subst_t2}), together with the gauge transformation (\ref{eq:gauge_t}), we have reduced the original Schr\"{o}dinger equation (\ref{eq:se_t}) to four independent tridiagonal matrices.

When diagonalizing any of the four matrices with elements given by Eqs.~(\ref{eq:T_sup}) - (\ref{eq:T_sub}), each of which pertains  to one of the four irreducible representations, we make use of the special properties of tridiagonal matrices, see e.g., Refs.~\cite{Fallat2011m, Veselic1979k}.
In particular, a tridiagonal matrix of dimension $M$,
\begin{eqnarray}\label{eq:tri}
D = \left( \begin{array}{ccccc}
a_0    & b_1 & 0      & \ldots &   0     \\
c_1    & a_1 & b_2    & 0      & \vdots  \\
0      & c_2 & a_2    & \ddots &         \\
\vdots &     & \ddots & \ddots & b_{M-1} \\
 0 & \ldots &  & c_{M-1} & a_{M-1}
\end{array} \right),
\end{eqnarray}
cannot be broken into block matrices if both $b_i\neq 0$ and $c_i \neq 0$. 
However, if there is an $N$ for which $b_N =0$ or $c_N=0$, the matrix $D$ can be broken into two tridiagonal matrices: a matrix, $D_1$, of dimension $N \times N$, and another  matrix, $D_2$, of size $(M-N)\times(M-N)$, with
\begin{equation}
\label{eq:sigma}
\sigma(D) = \sigma(D_1) \cup \sigma(D_2)
\end{equation}
where $\sigma(D)$ is the spectrum (set of eigenvalues) of matrix D.

For example, when $c_2=0$, we are left with the following block structure
\begin{eqnarray}\label{eq:tri0}
D = \left( \begin{array}{cc|ccc}
a_0    & b_1 & 0      & \ldots &   0     \\
c_1    & a_1 & b_2    & 0      & \vdots  \\
\hline
0      & \mathbf{0} & a_2    & \ddots &         \\
\vdots &     & \ddots & \ddots & b_{M-1} \\
 0 & \ldots &  & c_{M-1} & a_{M-1}
\end{array} \right),
\end{eqnarray}
indicated by the vertical and horizontal lines, in which case the eigenproperties can be calculated separately for the upper left $2\times 2$ and for the lower right $(M-2)\times(M-2)$ blocks. 
However, in the case of the tridiagonal matrices (\ref{eq:T_sup})-(\ref{eq:T_sub}) representing the symmetry-adapted Schr\"{o}dinger operator $T_{t,\kappa}^{(\Gamma_t)}$, the vanishing $c_N=0$ implies that only the upper left $N\times N$ block can be diagonalised explicitly.
Note that this is independent of the value of $b_N$. 
The lower right block which is of infinite dimension can be also diagonalized but cannot be computed explicitly.

Indeed, the subdiagonal elements given by Eq.~(\ref{eq:T_sub}) contain a single zero for each of the four symmetry-adapted matrices $T_{t,\kappa}^{(\Gamma_t)}$ for positive integer values of the topological index $\kappa$. 
For odd values of $\kappa$, the zeros of $T_{t,\kappa}^{(A_1)}$ occur at $\ell=(\kappa+1)/2$ and at $\ell=(\kappa-1)/2$ for $T_{t,\kappa}^{(A_2)}$ (however for $\kappa\ge 3$ only). 
As a result, the eigenproperties of $T_{t,\kappa}^{(A_1)}$ and $T_{t,\kappa}^{(A_2)}$ can be obtained analytically for the upper left blocks whose dimensions are
\begin{eqnarray}
N_{t,\kappa}^{(A_1)} &=& \frac{\kappa+1}{2} \nonumber \\
N_{t,\kappa}^{(A_2)} &=& \frac{\kappa-1}{2} \label{eq:N_t_A} \,.
\end{eqnarray} 
For even values of $\kappa$, the zeros occur at $\ell=\kappa/2$ for both $T_{t,\kappa}^{(B_1)}$ and $T_{t,\kappa}^{(B_2)}$ and the dimensions of the upper left blocks are 
\begin{eqnarray}
N_{t,\kappa}^{(B_1)} &=& \frac{\kappa}{2}  \nonumber \\
N_{t,\kappa}^{(B_2)} &=& \frac{\kappa}{2}  \label{eq:N_t_B} \,.
\end{eqnarray} 
None of the four finite-dimensional (upper left) blocks can be broken into smaller blocks, as their sub- and super-diagonal elements are all nonzero (in fact, positive). 
The infinite-dimensional (lower right) blocks cannot be broken into smaller blocks for the same reason (all superdiagonal entries are positive whereas all subdiagonal entries are negative).
Examples of the finite-dimensional matrices are presented in Section~\ref{sec:solve_ex} and used for calculating the eigenproperties of the quantum planar pendulum. 

The above provides a compelling explanation for the previously found conditionally quasi-exact solvability (C-QES) of the planar pendulum problem: 
If and only if $\kappa$ is an odd/even positive integer can the tridiagonal matrices, Eqs.~(\ref{eq:T_sup}) - (\ref{eq:T_sub}), corresponding to the $A_{1,2}/B_{1,2}$ irreducible representations,  be broken into finite-dimensional matrices (upper left) and infinite-dimensional remainders (lower right), whereby the finite-dimensional matrices can be diagonalized, at least in principal, analytically, with solutions that are periodic/antiperiodic in $2\pi$. 

Note that for $\kappa$ non-integer, there will be no zeros on the sub- or super-diagonals of the matrices (\ref{eq:T_sup}) - (\ref{eq:T_sub}), which will thus be of infinite dimension and, therefore, not amenable to analytic diagonalization.

The finite dimensions of the upper left block matrices, Eqs.~(\ref{eq:N_t_A}) and (\ref{eq:N_t_B}) for odd and even $\kappa$, respectively, determine the number of analytic solutions of the Schr\"odinger equation (\ref{eq:se_t}). 
Since $N_{t,\kappa}^{(A_1)}+N_{t,\kappa}^{(A_2)}=\kappa$ for odd $\kappa$ and $N_{t,\kappa}^{(B_1)}+N_{t,\kappa}^{(B_2)}=\kappa$ for even $\kappa$,  we see that the number of analytic solutions is in any case equal to the topological index $\kappa$ itself. 
Thus, for a given $\kappa$, a finite number of analytic solutions is obtained and, therefore, the planar pendulum problem is QES.
We note that in practice the number of analytic eigenvalues and eigenfunctions is limited to $N_{t,\kappa}^{(\Gamma_t)} \le 4$. For more, see also the following Section~\ref{sec:solve_ex}.

Ultimately, our method starting from the identification of the four finite irreducible representations $\Gamma_t\in \left\{A_1,B_1,B_2,A_2\right\}$ of Hamiltonian (\ref{eq:se_t}) is equivalent to building four $N_{t,\kappa}^{(\Gamma_t)}$-dimensional monomial subspaces each of which is invariant under the action of the corresponding symmetry-adapted  operator $T_{t,k}^{(\Gamma_t)}$, see Eq.~(\ref{eq:gauge_t}).
This circumstance suggests that our method is related to Lie algebraic methods, see, e.g., Refs. \cite{Gomez2005d, Leon2014a}.
We note that for $\kappa$ non-integer, there are no invariant subspaces, in which case the infinite tridiagonal matrices cannot be reduced. 
However, they can be diagonalised numerically. 

\subsection{Razavy potential}
\label{sec:solve_h} 
In analogy with the procedure introduced in Section \ref{sec:solve_t} for the planar pendulum,  we make use of the substitution 
\begin{equation}
\label{eq:subst_h1}
\psi_h(\theta) = f_h(\theta)\exp(\beta\cosh\theta)
\end{equation}
where $\beta<0$ is used to ensure a correct asymptotic behavior. 
This substitution serves to recast the original Schr\"{o}dinger equation (\ref{eq:se_h}) for the Razavy system as 
\begin{equation}
\label{eq:ince_h}
-\frac{d^2f_{h}(x)}{dx^2}-2 \beta\sinh x \frac{df_{h}(x)}{dx} + \left[\beta^2 + \beta (\kappa-1)\cosh x\right] f_{h}(x)=E_{h} f_{h}(x) 
\end{equation}
which is a hyperbolic analog of the Ince equation (\ref{eq:ince_t}); we note that Eq. (\ref{eq:ince_h}) can be obtained directly from Eq. (\ref{eq:ince_t})  by an anti-isospectral transform: $x\mapsto -i\theta$, $f_h(x) \mapsto f_t(\theta)$, $E_h=-E_t$. 

With the further substitution
\begin{equation}
\label{eq:subst_h2}
u \equiv \cosh\frac{x}{2} 
\end{equation}
the hyperbolic Ince equation (\ref{eq:ince_h}) can be written as 
\begin{eqnarray}
\label{eq:T_h}
T_{h,\kappa} \phi_{h,\kappa}&\equiv &\frac{1}{4}(1-u^{2})\frac{d^2\phi_{h,\kappa}}{du^2} + \left(2\beta u -2\beta u^3-\frac{u}{4}\right)\frac{d\phi_{h,\kappa}}{du}+ \left[(2u^2-1)(\kappa-1)\beta+\beta^2\right]\phi_{h,\kappa} \nonumber \\
&=& E_{h,\kappa}\phi_{h,\kappa} \,.
\end{eqnarray}
Here, $T_{h,\kappa}$ is the Schr\"{odinger} operator for the Razavy system and $\phi_{h,\kappa}(u)$ is equivalent to $f_h(x)$ for a given value of $\kappa$. 
Note that by virtue of the last substitution, all hyperbolic functions have been eliminated.
Applying transformations (\ref{eq:subst_h1}) and (\ref{eq:subst_h2}) to the two hyperbolic seed functions, cf. Table~\ref{tab:seed} and Figure~\ref{fig:seed}, yields
\begin{eqnarray}
\label{eq:seed_h}
\phi_{h,1}^{(A')}(u) &=& 1 \nonumber \\
\phi_{h,2}^{(A'')}(u) &=& \pm \sqrt{u^2-1}\,
\end{eqnarray}
which are identical with the expressions (\ref{eq:seed_t}) for the seed functions $\psi_{t,1}^{(A_1)}$ and $\psi_{t,2}^{(B_2)}$ of the pendular system. 
This identity results from the correlation between the four irreducible representations $\Gamma_t$ of the $C_{2v}$ group with the two irreducible representations $\Gamma_h$ of its $C_i$ subgroup, cf. Tab.~\ref{tab:seed} and Fig.~\ref{fig:seed}. 
Again, these lowest-order eigenfunctions that transform according to the irreducible representations of the $C_i$ point group can be used to symmetry-adapt the Schr\"{o}dinger operator $T_{h,\kappa}$, see Eq.~(\ref{eq:T_h}), to the symmetry of the Razavy system via the following gauge transformation, 
\begin{equation}
\label{eq:gauge_h}
T_{h,\kappa}^{(\Gamma_h)} \equiv \frac{1}{\phi_{h,\kappa}^{(\Gamma_h)}}T_{h,\kappa}\phi_{h,\kappa}^{(\Gamma_h)}
\end{equation}
where $\Gamma_h\in \left\{A',A''\right\}$ with $\kappa\in\{1,2\}$. 
Like in the trigonometric case,  the symmetry-adapted Schr\"{o}dinger operators $T_{h,\kappa}^{(\Gamma_h)}$ in the hyperbolic case has the same spectrum as the original operator of $T_{h,\kappa}$ \cite{kamran1990,Glopez1994d}. 

In order to obtain explicit matrix representations of the $T_{h,\kappa}^{(\Gamma_h)}$ operators, we make use of a basis set of monomials in $u$
\begin{equation}
\label{eq:basis_h}
\left\{1,u,u^2,\ldots\right\} \,.
\end{equation}
In contrast to the trigonometric case, the hyperbolic basis set is comprised of  both even- and odd-order monomials, as $u=\cosh(x/2)$ is totally symmetric (i.e., has even parity and pertains to  the $A'$ irreducible representation) with respect to the symmetry operations of the Razavy system as given by Eq. (\ref{eq:map_Ci}) and listed in Table \ref{tab:irreps} and thus not affecting the symmetry of the $T_{h,\kappa}^{(\Gamma)}$ operators.

Using the basis set of Eq.~(\ref{eq:basis_h}), the non-zero matrix elements of the symmetry-adapted Schr\"{o}dinger operators for the Razavy system $T_{h,\kappa}^{(\Gamma_h)}$ can be expressed in terms of the corresponding operators for the planar pendulum $T_{t,\kappa}^{(\Gamma_t)}$,
\begin{eqnarray}
&\langle u^{2\ell  } | T_{h,\kappa}^{(A' )} | u^{2\ell'  } \rangle &= \langle u^{2\ell} | T_{t,\kappa}^{(A_1)} | u^{2\ell'} \rangle \nonumber \\
&\langle u^{2\ell+1} | T_{h,\kappa}^{(A' )} | u^{2\ell'+1} \rangle &= \langle u^{2\ell} | T_{t,\kappa}^{(B_1)} | u^{2\ell'} \rangle \nonumber \\
&\langle u^{2\ell  } | T_{h,\kappa}^{(A'')} | u^{2\ell'  } \rangle &= \langle u^{2\ell} | T_{t,\kappa}^{(B_2)} | u^{2\ell'} \rangle \nonumber \\
&\langle u^{2\ell+1} | T_{h,\kappa}^{(A'')} | u^{2\ell'+1} \rangle &= \langle u^{2\ell} | T_{t,\kappa}^{(A_2)} | u^{2\ell'} \rangle \label{eq:T_h_mat}
\end{eqnarray}
with  $\ell'=\ell$ for the main diagonal and $\ell'=\ell\pm 2$ for sub- and super-diagonals.
The first and third identities are true by definition, because the seed functions for the $A'$ and $A''$ irreducible representations of the hyperbolic system ($C_i$) have been chosen to be identical with the seed functions for the $A_1$ and $B_2$ irreducible representations of the trigonometric system ($C_{2v}$).
The second identity reflects the fact that the $A_1$ and $B_1$ seed functions of the pendulum differ by one power of $u$, i.e., $\phi_{t,\kappa}^{(B_1)}(u)=u\phi_{t,\kappa}^{(A_1)}(u)=u\phi_{h,\kappa}^{(A')}(u)$. 
The same holds for the $B_2$ and $A_2$ seed functions, $\phi_{t,\kappa}^{(A_2)}(u)=u\phi_{t,\kappa}^{(B_2)}(u)=iu\phi_{h,\kappa}^{(A'')}(u)$, which leads to the fourth identity of Eq.~(\ref{eq:T_h_mat}).

Note that all other matrix elements of $T_{h,\kappa}^{(\Gamma_h)}$, i.e., those coupling even with odd powers of $u$, vanish. 
This is because of the structure of the Schr\"{o}dinger operator $T_{h,\kappa}$, see Eq.~(\ref{eq:T_h}).
Since $T_{h,\kappa}$ leaves both the space of even-ordered and odd-ordered monomials invariant, we also end up with  four matrices, in complete analogy to the four tridiagonal matrices (\ref{eq:T_sup}-\ref{eq:T_sub}) occuring for the trigonometric case, even though the reduced $C_i$ symmetry of the hyperbolic problem allows for a decomposition of the original Hamiltonian matrix into two blocks only ($A'$ and $A''$).
Hence, from here on we write $T_\kappa^{(\Gamma)}=T_{t,\kappa}^{(\Gamma)}=T_{h,\kappa}^{(\Gamma)}$, i.e., we drop the subscripts {\it t} and {\it h}, and use
the $\Gamma\in\{A_1,B_1,B_2,A_2\}$ labelling, originally introduced for the trigonometric system, for the hyperbolic system as well.
The same applies for the dimensions of the corresponding matrices $N_\kappa^{(\Gamma)}$ defined in Eqs.~(\ref{eq:N_t_A}) and (\ref{eq:N_t_B}).

In summary, as implied by the equality of the matrix representations of the respective Schr\"{o}dinger operators, Eq. (\ref{eq:T_h_mat}), the hyperbolic Razavy system is, like the planar pendulum, a C-QES system, i.e., analytic solutions can only be found under the condition that the topological index $\kappa$ be an integer.
At the same time, the Razavy system is also QES, i.e., only a finite number of analytic solutions exist, and this number is given by the value of $\kappa$. 
Hence, all the analytic eigenenergies (for integer $\kappa$) of the planar pendulum are also the eigenenergies of the Razavy system, however, with an opposite sign as required by the anti-isospectrality condition (\ref{eq:ais}),
\begin{equation}
\label{eq:ais_E}
\left(E_{h,\kappa}^{(\Gamma)}\right)_n=-\left(E_{t,\kappa}^{(\Gamma)}\right)_{N_\kappa^{(\Gamma)}-n-1}
\end{equation}
with quantum numbers $0\le n \le N_\kappa^{(\Gamma)}-1$ the ordering of which is reversed within each irreducible representation $\Gamma$.

\subsection{Sample calculations}
\label{sec:solve_ex}
In this section, we delve into the details of extracting analytic eigenproperties of the planar pendulum from the general theory presented above.  We begin by writing out explicitly the finite-dimensional tridiagonal block matrices representing the symmetry-adapted Schr\"{o}dinger operators, Eq. (\ref{eq:gauge_t}) in the monomial basis (\ref{eq:basis_t}), whereby we make use of the matrix elements given by Eqs.~(\ref{eq:T_sup}) - (\ref{eq:T_sub}) as well as of the blocks' dimensions, given by Eqs. (\ref{eq:N_t_A}) and (\ref{eq:N_t_B})
\begin{eqnarray}
\label{eq:T_t_A1}
T_{\kappa}^{(A_1)} &=& \left( \begin{array}{cccc}
\beta^{2} -(\kappa-1)\beta & \frac{1}{2} & \ldots & 0 \\
 2(\kappa-1)\beta & \beta^{2}-1+4\beta-(\kappa-1)\beta & \ddots & \vdots \\
\vdots & \ddots & \ddots & \frac{(\kappa-1)^2}{4}-\frac{\kappa-1}{4}\\
 0 &  0 & 4\beta  & \beta^{2}-\frac{(\kappa-1)^2}{4} + (\kappa-1)\beta
\end{array} \right)\\
\label{eq:T_t_B1}
T_{\kappa}^{(B_1)} &=& \left( \begin{array}{cccc}
\beta^{2} -\frac{1}{4} - (\kappa-3)\beta   & \frac{3}{2}& \ldots & 0 \\
 2(\kappa-2)\beta  & \beta^2-\frac{9}{4}+4\beta-(\kappa-3)\beta & \ddots & \vdots \\
\vdots & \ddots & \ddots & \frac{(\kappa-2)^2}{4}+\frac{\kappa-2}{4} \\
 0 &  0 & 4\beta & \beta^{2} -\frac{(\kappa-1)^2}{4}+ (\kappa-1)\beta
\end{array} \right)\\
\label{eq:T_t_B2}
T_{\kappa}^{(B_2)} &=& \left( \begin{array}{cccc}
\beta^{2} -\frac{1}{4} -(\kappa-1)\beta & \frac{1}{2} & \ldots & 0 \\
2(\kappa-2)\beta & \beta^{2}-\frac{9}{4}+4\beta-(\kappa-1)\beta& \ddots & \vdots \\
\vdots & \ddots & \ddots &\frac{(\kappa-2)^2}{4}-\frac{\kappa-2}{4}\\
 0 &  0 &  4\beta & \beta^{2} -\frac{(\kappa-1)^2}{4} + (\kappa-3)\beta
\end{array} \right)\\
\label{eq:T_t_A2}
T_{\kappa}^{(A_2)} &=& \left( \begin{array}{cccc}
\beta^{2} -1 -(\kappa-3)\beta  &  \frac{3}{2} & \ldots & 0 \\
2(\kappa-3)\beta  & \beta^{2} -4 + 4\beta -(\kappa-3)\beta & \ddots & \vdots \\
\vdots & \ddots & \ddots & \frac{(\kappa-3)^2}{4}+\frac{\kappa-3}{4}  \\
 0 &  0 & 4\beta & \beta^{2} - \frac{(\kappa-1)^2}{4} + (\kappa-3)\beta
\end{array} \right)
\end{eqnarray}

Note again that these matrices are the same for the trigonometric and hyperbolic system, $T_{\kappa}^{(\Gamma)}=T_{t,\kappa}^{(\Gamma)}=T_{h,\kappa}^{(\Gamma)}$ where the four irreducible representations of the former system are also used for the latter one, see above.
As before, the $A_{1,2}$/$B_{1,2}$ representations pertain, respectively, to odd/even $\kappa$.
Analytic eigenenergies  $E_{t,\kappa}^{(\Gamma)}$ of the pendulum's Schr\"{o}dinger equation (\ref{eq:se_t}) and $E_{h,\kappa}^{(\Gamma)}$ of the Razavy equation  (\ref{eq:se_h})  are then obtained as the negative or positive eigenvalues of these four matrices, repectively, see also the definition of the $T_{t,\kappa}$ operator in Eq.~(\ref{eq:T_t}), and $T_{h,\kappa}$ operator in Eq.~(\ref{eq:T_h}).

Because in general only matrices up to dimension four can be diagonalised analytically, it follows from Eqs.~(\ref{eq:N_t_A}),(\ref{eq:N_t_B}) that all $A_1$ and $A_2$ solutions for $\kappa$ odd up to $\kappa=7$, as well as four $A_2$ solutions for $\kappa=9$ could be obtained.
For $\kappa$ even, all $B_1$ and $B_2$ solutions up to $\kappa=8$ could be obtained. 
This gives a total of forty analytic solutions which we obtained using computer algebra systems (both Symbolic Toolbox of Matlab and Mathematica).
The eigenenergies of the twenty four lowest states ($N_{\kappa}^{(\Gamma)}\le 3$) are listed in Table~\ref{tab:ana_energies}; the remaining sixteen solutions ($N_{\kappa}^{(\Gamma)}=4$) are available from the authors upon request.
The eigenenergies for the particular choice of $\beta=-5$ are shown in bold face in Table~\ref{tab:num_pendulum} for the trigonometric system and in Table~\ref{tab:num_razavy} for the Razavy system. 
An inspection of Tabs.~\ref{tab:ana_energies},\ref{tab:num_pendulum},\ref{tab:num_razavy} reveals that the eigenenergies derived from the different irreducible representations $A_1, A_2$ or $B_1, B_2$ for odd and even $\kappa$, respectively, are interleaved and form the spectrum \cite{Hemery2010a}.

The analytic eigenfunctions corresponding to the above analytic eigenenergies  can be obtained in analytic form as products of gauge factors and polynomials in $u$

\begin{eqnarray}
\label{eq:eig_t_AB}
\left(\psi_{t,\kappa}^{(A_1)}\right)_n \propto e^{\beta\cos\theta} \sum_{\ell=0}^{(\kappa-1)/2} \left( v_{\kappa}^{(A_1)}\right)_{n,\ell} \cos^{2\ell}  \frac{\theta}{2} \nonumber \\
\left(\psi_{t,\kappa}^{(B_1)}\right)_n \propto e^{\beta\cos\theta} \cos\frac{\theta}{2} \sum_{\ell=0}^{(\kappa-2)/2} \left( v_{\kappa}^{(B_1)}\right)_{n,\ell} \cos^{2\ell}\frac{\theta}{2} \nonumber \\
\left(\psi_{t,\kappa}^{(B_2)}\right)_n \propto e^{\beta\cos\theta} \sin\frac{\theta}{2} \sum_{\ell=0}^{(\kappa-2)/2} \left( v_{\kappa}^{(B_2)}\right)_{n,\ell} \cos^{2\ell}  \frac{\theta}{2} \nonumber \\
\left(\psi_{t,\kappa}^{(A_2)}\right)_n \propto e^{\beta\cos\theta} \sin\theta \sum_{\ell=0}^{(\kappa-3)/2} \left( v_{\kappa}^{(A_2)}\right)_{n,\ell} \cos^{2\ell}\frac{\theta}{2}
\end{eqnarray}
Note that for the eigenfunctions $\psi_{h,\kappa}^{(\Gamma)}$ of the Razavy case, all trigonometric functions should be replaced by their hyperbolic counterparts.
The leading term is the von Mises function, $\exp(\beta\cos\theta)$, for the trigonometric system, or its equivalent, $\exp(\beta\cosh x)$, for the hyperbolic system.
The second term is a seed function pertaining to one of the four irreducible representations in question, and $(v_{\kappa}^{(\Gamma)})_{n,\ell}$ are the coefficients of the monomials $u^{2\ell}$. 
These coefficients are the eigenvectors of the matrices $T_{\kappa}^{(\Gamma)}$ given in Eqs.~(\ref{eq:T_t_A1}) -- (\ref{eq:T_t_A2}).

\subsubsection{The case of $N_{\kappa}^{(\Gamma)}=1$}

According to Eqs. (\ref{eq:N_t_A}) and (\ref{eq:N_t_B}), the value of $N_{\kappa}^{(\Gamma)}=1$ admits  the values of $\kappa=1,2,3$. For $\kappa=1$,  only the $A_1$ representation furnishes $N_{\kappa}^{(\Gamma)} =N_{1}^{(A_1)}=1$, cf. Eq. (\ref{eq:N_t_A}). In this case matrix (\ref{eq:T_t_A1}) trivially reduces to its upper left element which equals the negative of the corresponding $A_1$ eigenenergy,
\begin{equation}
\label{eq:E_k1_A1}
\left(E_{t,1}^{(A_1)}\right)_0 = -\left(E_{h,1}^{(A_1)}\right)_0= -\beta^2 \,
\end{equation}

For $\kappa=2$, Eq.~(\ref{eq:N_t_B}) implies that the problem reduces to two one-dimensional problems, with  $N_{2}^{(B_1)}=N_{2}^{(B_2)}=1$.
The corresponding eigenenergies of the $B_1$ and $B_2$ states are, respectively, the negative of the upper left elements of matrices (\ref{eq:T_t_B1}) and (\ref{eq:T_t_B2}),
\begin{eqnarray}
\label{eq:E_k1_B12}
\left(E_{t,2}^{(B_1)}\right)_0 &=& -\left(E_{h,2}^{(B_1)}\right)_0 = -\beta^2-\beta +\frac{1}{4} \nonumber \\
\left(E_{t,2}^{(B_2)}\right)_0 &=& -\left(E_{h,2}^{(B_2)}\right)_0 = -\beta^2+\beta +\frac{1}{4} \, 
\end{eqnarray}

For $\kappa=3$, only the  $A_2$ representation furnishes $N_{1}^{(A_2)}=1$, cf. Eq.~(\ref{eq:N_t_A}), whose eigenenergy is obtained from the upper left element of matrix (\ref{eq:T_t_A2})
\begin{equation}
\label{eq:E_k3_A2}
\left(E_{t,3}^{(A_2)}\right)_0 = -\left(E_{h,3}^{(A_2)}\right)_0 = -\beta^2+1 \, 
\end{equation}

The corresponding eigenvector matrices $v_\kappa^{(\Gamma)}$ with $N_{\kappa}^{(\Gamma)}=1$ simply reduce to a scalar that can be plugged into Eq. (\ref{eq:eig_t_AB}) to yield the wavefunctions.
As can be see in Table \ref{tab:seed},  the four eigenenergies and eigenfunctions for $N_{\kappa}^{(\Gamma)} =1$ reduce to those for the seed functions, cf. Section~\ref{sec:solve_t} above.  
Note that these four states were already known for the pendular case from our previous work,  where they were obtained via supersymmetry (SUSY QM) \cite{Schmidt2014b} and for the hyperbolic case from Razavy's original work \cite{Razavy1980m}. Note that for both cases this $A_2$ state is the first excited state for $\kappa=3$.

\subsubsection{The case of $N_{\kappa}^{(\Gamma)}=2,3$}
In addition to the $A_2$ state for $\kappa=3$ mentioned above, there are also two totally symmetric solutions with $N_{3}^{(A_1)}=2$. For this case, matrix  (\ref{eq:T_t_A1}) simplifies to
\begin{eqnarray}
T_{3}^{(A_1)} = \left( \begin{array}{cc}
\beta^2-2\beta & \frac{1}{2} \\
4\beta & \beta^2+2\beta-1 
\end{array} \right),
\end{eqnarray}
whose eigenvalues give the eigenenergies 
\begin{equation}
\label{eq:E_k3_A1}
\left(E_{t,3}^{(A_1)}\right)_{0/1} = -\left(E_{h,3}^{(A_1)}\right)_{1/0}  = - \beta^2 \mp \frac{1}{2}\sqrt{16\beta^2 + 1} + \frac{1}{2}  
\end{equation}
The corresponding wavefunctions for the trigonometric case
\begin{equation}
\left(\psi_{t,3}^{(A_1)}\right)_{0/1} \propto e^{\beta\cos\theta}\left(\frac{1\pm\sqrt{16\beta^2+1}-4\beta}{8\beta} + \cos^2\frac{\theta}{2}\right) 
\end{equation}
are of even parity and $2\pi$-periodic, as required for the totally symmetric $A_1$ representation. 
For the eigenfunctions $\psi_{h,3}^{(A_1)}$ of the hyperbolic system the cos functions have to be replaced by their  hyperbolic counterpart.

We skip the case of $\kappa=4$ where there are a $B_1$ and a $B_2$ representations, each of them two-dimensional, and continue with the case of $\kappa=5$.
In accordance with Eq.~(\ref{eq:N_t_A}), $N_{5}^{(A_1)}=3$ and $N_{5}^{(A_2)}=2$.
The three-dimensional $A_1$ representation obtained from matrix (\ref{eq:T_t_A1}) yields the following tridiagonal matrix   
\begin{equation}
T_{5}^{(A_1)} = \left( \begin{array}{ccc}
\beta^2-4\beta & \frac{1}{2} & 0  \\
8\beta         & \beta^2-1   & 3  \\
0              & 4\beta      & \beta^2+4\beta-4 
\end{array} \right)
\end{equation}
and the two-dimensional $A_2$ representations from matrix (\ref{eq:T_t_A2}) yields   
\begin{equation}
T_{5}^{(A_2)} = \left( \begin{array}{cc}
\beta^2-2\beta-1 & \frac{3}{2} \\
4\beta           & \beta^2+2\beta-4 
\end{array} \right) \,.
\end{equation}
Therefore, we only need to diagonalize these matrices, instead of a $5 \times 5$ Hamiltonian, which is not possible to do analytically in general. 
Analytic expressions for the eigenvalues are listed in Table~\ref{tab:ana_energies} and the numeric expressions for the specific choice of $\beta=-5$  in Tabs.~\ref{tab:num_pendulum}, \ref{tab:num_razavy}.
We note that the eigenvalues of the two symmetries are interleaved.

The corresponding wavefunctions can be calculated from Eq. (\ref{eq:eig_t_AB}). For the $A_1$ symmetry, we obtain  
\begin{eqnarray}
\label{eq:psi5A1}
\left(\psi_{t,5}^{(A_1)}\right)_0 &\propto& e^{\beta \cos(\theta)} \left((v_{5}^{(A_1)})_{0,0} + (v_{5}^{(A_1)})_{0,1} \cos^2\frac{\theta}{2} + (v_{5}^{(A_1)})_{0,2}\cos^4\frac{\theta}{2} \right)	\nonumber \\
\left(\psi_{t,5}^{(A_1)}\right)_1 &\propto& e^{\beta \cos(\theta)} \left((v_{5}^{(A_1)})_{1,0} + (v_{5}^{(A_1)})_{1,1} \cos^2\frac{\theta}{2} + (v_{5}^{(A_1)})_{1,2}\cos^4\frac{\theta}{2} \right) \nonumber \\
\left(\psi_{t,5}^{(A_1)}\right)_2 &\propto& e^{\beta \cos(\theta)} \left((v_{5}^{(A_1)})_{2,0} + (v_{5}^{(A_1)})_{2,1} \cos^2\frac{\theta}{2} + (v_{5}^{(A_1)})_{2,2}\cos^4\frac{\theta}{2} \right)
\end{eqnarray}
and for the $A_2$ symmetry these are
\begin{eqnarray}
\label{eq:psi5A2}
\left(\psi_{t,5}^{(A_2)}\right)_0 &\propto& \sin \theta e^{\beta \cos(\theta)} \left((v_{5}^{(A_2)})_{0,0} + (v_{5}^{(A_2)})_{0,1} \cos^2\frac{\theta}{2}\right) \nonumber \\
\left(\psi_{t,5}^{(A_2)}\right)_1 &\propto& \sin \theta e^{\beta \cos(\theta)} \left((v_{5}^{(A_2)})_{1,0} + (v_{5}^{(A_2)})_{1,1} \cos^2\frac{\theta}{2}\right)
\end{eqnarray}
with the requisite eigenvectors given as rows (the numbering of which starts from 0) of the following matrices
\begin{equation}
v_{5}^{(A_1)} = \left( 
\begin{array}{ccc}
1.15537 & -2.15340 & 1 \\
0.02286 & -1.05075 & 1 \\
0.00177 & -0.14584 & 1 
\end{array} 
\right),\quad
v_{5}^{(A_2)} = \left( 
\begin{array}{cc}
-1.08059 & 1 \\
-0.06941 & 1 
\end{array} 
\right)
\end{equation}
for $\beta=-5$ as an example. Again, for the eigenfunctions $\psi_{h,5}^{(A_1)},\psi_{h,5}^{(A_2)}$ of the hyperbolic Razavy system all trigonometric functions have to be replaced by their hyperbolic counterparts.

\subsection{Discussion of limiting cases}
\label{sec:limits}
\subsubsection{The case of $|\beta|>\kappa/2$}
\begin{figure}
\includegraphics[width=17cm]{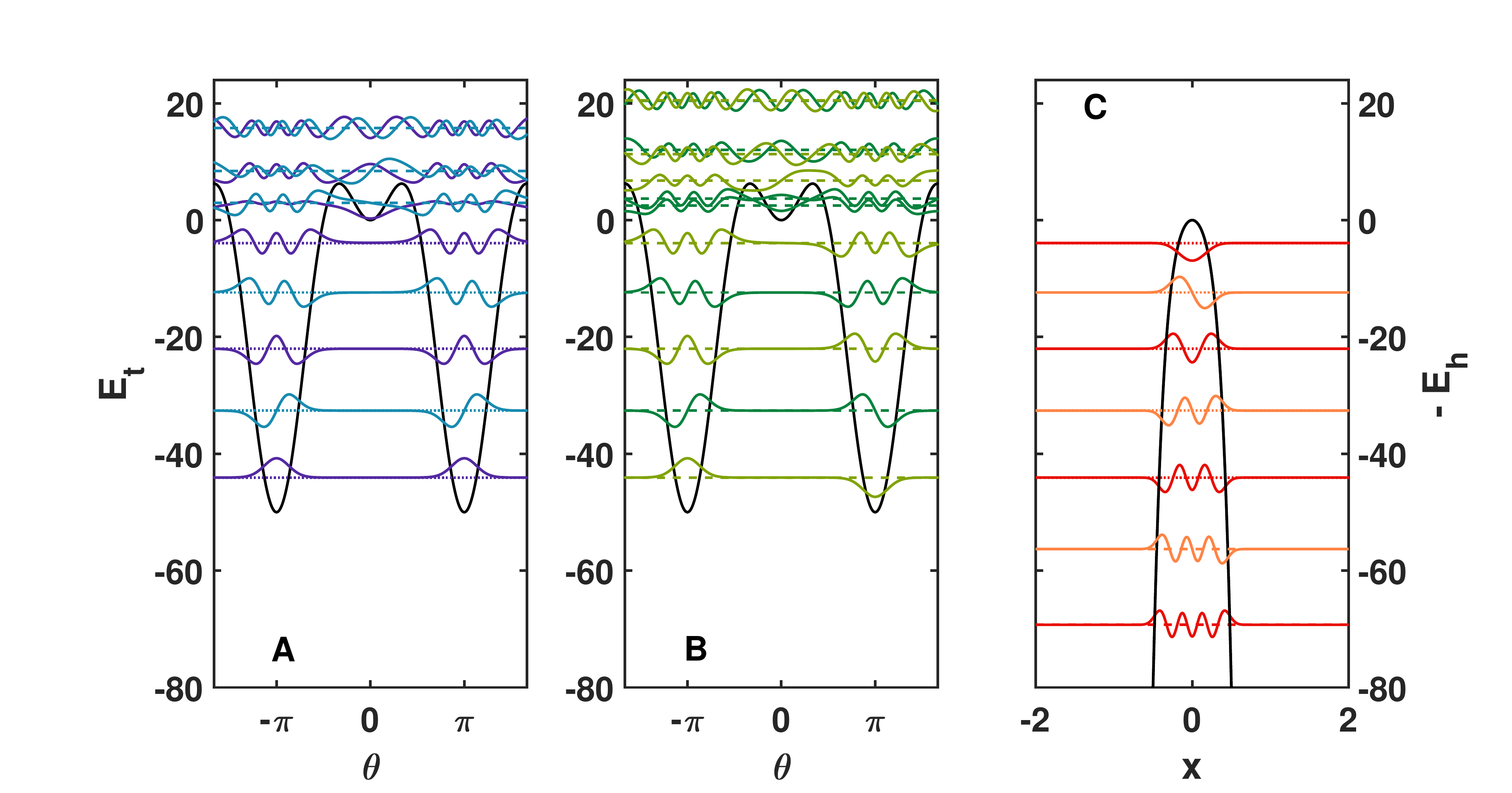}
\caption{Trigonometric $V_t$ (A,B) and inverted hyperbolic $-V_h$ (C) potentials for $\beta=-5$, with analytical (dotted lines) and numerical (dashed lines) eigenenergies and wavefunctions (full curves). For $\kappa=5$, the energies $E_t$ for periodic states (A) are anti-isospectral with Razavy energies $E_h$.}
\label{fig:k5_b500}
\end{figure}

As mentioned in Section~\ref{sec:ham_t} and \ref{sec:ham_h}, for the case of $|\beta|>\kappa/2$, or equivalently, $|\eta|<2\zeta$, the trigonometric potential is an asymmetric double well, whereas the hyperbolic potential has just  a single well.
This is illustrated for $\kappa=5$ in Figure~\ref{fig:k5_b500} where we show the eigenvalues, see also Tables~\ref{tab:num_pendulum} and \ref{tab:num_razavy} and eigenfunctions, see Eqs.~(\ref{eq:psi5A1}) and (\ref{eq:psi5A2}), for the value of $\beta=-5$.
As implied by the odd value of $\kappa$, the five analytic states for the trigonometric case are 2$\pi$-periodic ($A_{1,2}$).
These (single) energy levels are the lowest A states, and they are located below the potential's secondary (local) minimum, see panel A of  Figure~\ref{fig:k5_b500}.
The corresponding numerical solutions for  2$\pi$-antiperiodic states ($B_{1,2}$) are shown in panel B.
With the energy barrier, $(|\beta|+\kappa/2)^2$, given as the difference between global minima and maxima, being large, the tunnel splitting is very small, hardly visible on the scale of the figure for the example of $\beta=-5$.

A comparison with panel C of Figure~\ref{fig:k5_b500} reveals that the five analytic A-states are anti-isospectral with the five lowest states of the Razavy system, which is a single well for $\beta=-5$.
As noted in Section~\ref{sec:solve_symm}, the $A_1$ ($A_2$) states of the $C_{2v}$ group correlate with the $A'$ ($A''$) of the $C_i$ group. 

\begin{figure}
\includegraphics[width=17cm]{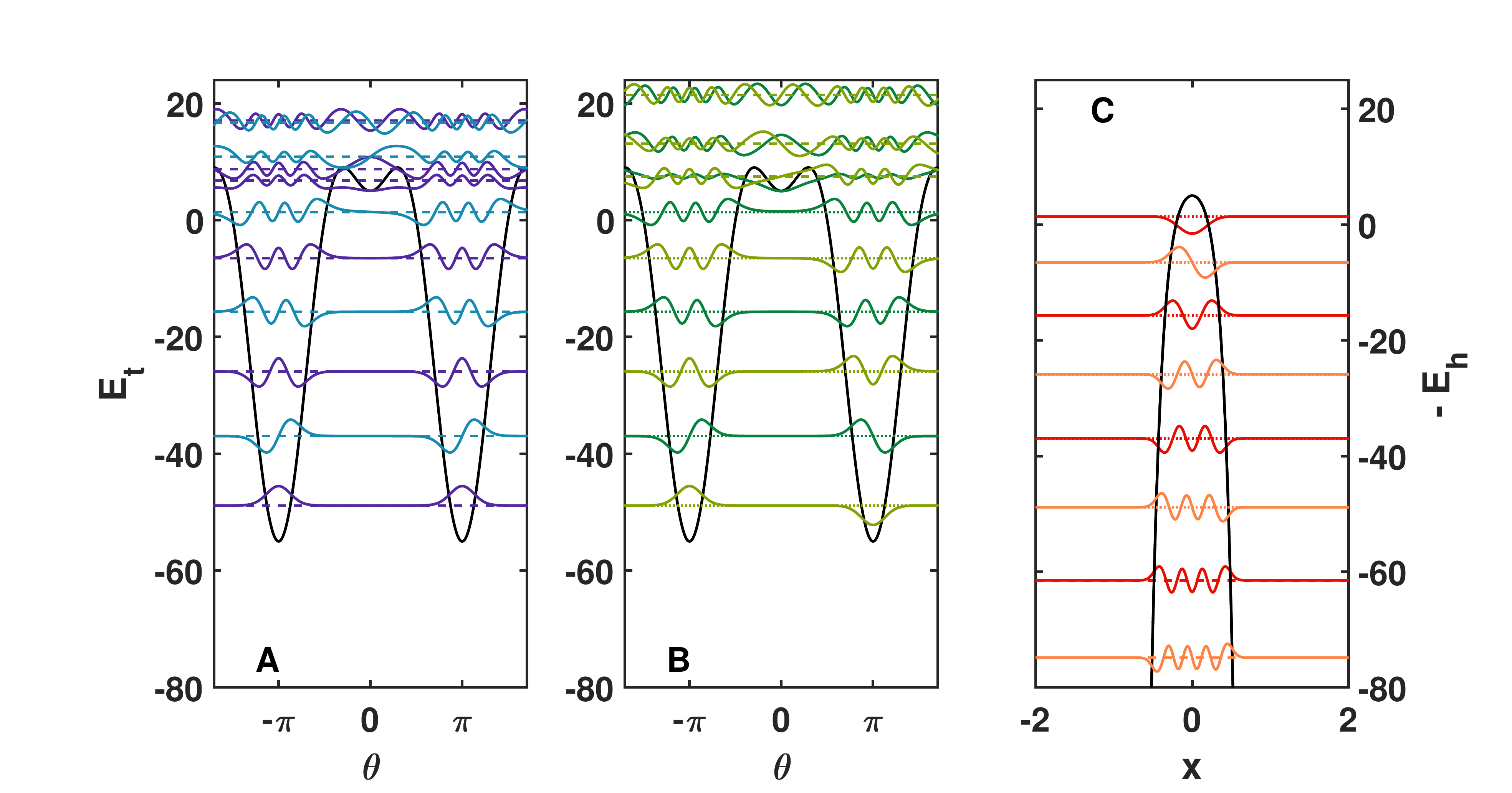}
\caption{Same as Fig.~\ref{fig:k5_b500} but for $\kappa=6$, where the energies $E_t$ for anti-periodic pendular states (B) are anti-isospectral to $E_h$.} 
\label{fig:k6_b500}
\end{figure}

Results for $\kappa=6$ are shown in Figure~\ref{fig:k6_b500} where the six analytic states for the trigonometric case are 2$\pi$-antiperiodic ($B_{1,2}$), see panel B of that figure.
Again, the corresponding numerical solutions for 2$\pi$-periodic states ($A_{1,2}$, see panel A) are separated from the $B$ energy levels by very small tunneling splittings. 
Because $\kappa$ is an even integer here, the B states are anti-isospectral with states of the hyperbolic system (see panel C) where now the $B_1$ ($B_2$) states of the $C_{2v}$ group correlate with the $A'$ ($A''$) of the $C_i$ group.

It can also be seen in Figures~\ref{fig:k5_b500} and \ref{fig:k6_b500} that the progressions of the analytic eigenenergies are qualitatively similar to those of a harmonic oscillator. 
In fact, in the limit of large $|\beta|$ the analytic energies given in Tab.~\ref{tab:ana_energies} converge to being equidistant, with a spacing of $2|\beta|$ centered around $-\beta^2$.

\subsubsection{The case of $|\beta|<\kappa/2$}
\begin{figure}
\includegraphics[width=17cm]{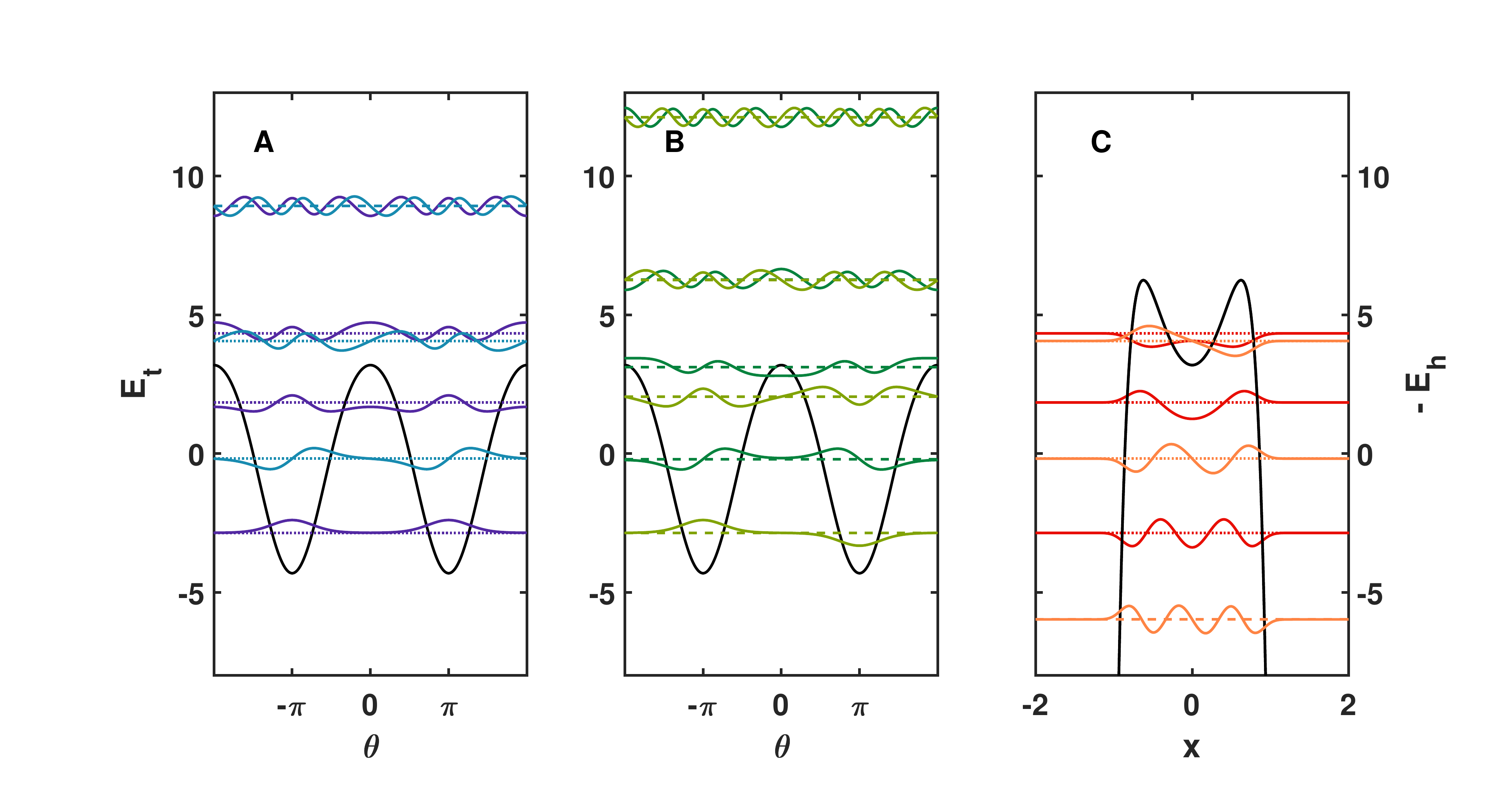}
\caption{Same as Fig.~\ref{fig:k5_b500} but for $\beta=-3/4$. For $\kappa=5$, the energies $E_t$ for periodic states (A) are anti-isospectral to $E_h$.}
\label{fig:k5_b075}
\end{figure}

\begin{figure}
\includegraphics[width=17cm]{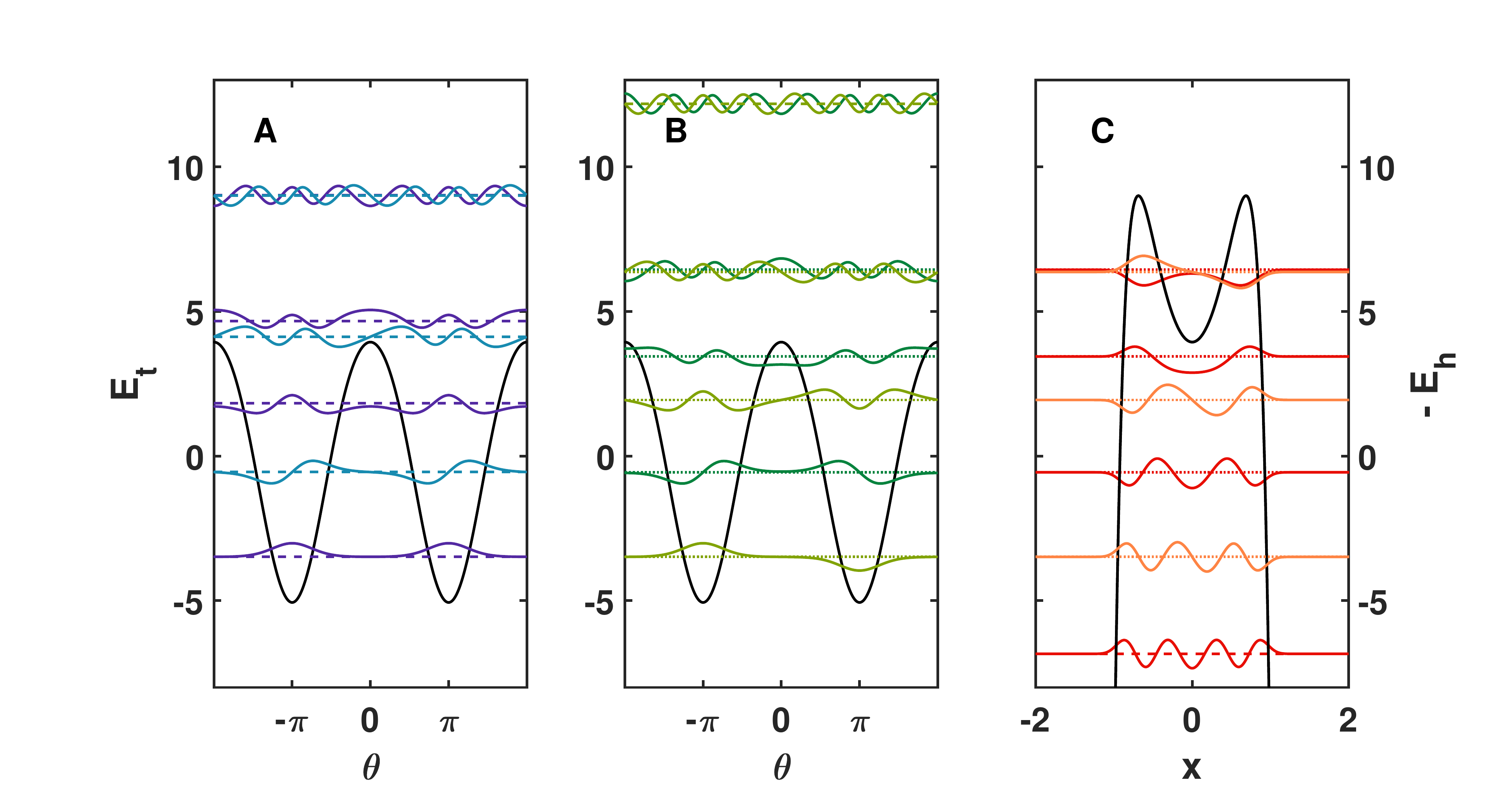}
\caption{Same as Fig.~\ref{fig:k5_b075} but for $\kappa=6$, where the energies $E_t$ for anti-periodic pendular states (B) are anti-isospectral to $E_h$.}
\label{fig:k6_b075}
\end{figure}

For the case of $|\beta|<\kappa/2$, or equivalently $|\eta|>2\zeta$, the trigonometric potential has a single well whereas the hyperbolic potential becomes a double well potential.
This is illustrated in Figures~\ref{fig:k5_b075} and \ref{fig:k6_b075} for $\beta=-3/4$, which display eigenenergies and eigenfunctions for $\kappa=5$ (or $\kappa=6$), respectively. Again the A (or B) energy levels are anti-isospectral with the eigenenergies of the Razavy system shown in panels C of the two figures.
For $\kappa=5$ ($\kappa=6$), there are three A states (four B states) below the maximum of the trigonometric potential or above the barrier of the hyperbolic potential.
These states are again essentially like harmonic oscillator states, but slightly affected by tunneling in some cases. 
The remaining two analytic $A$ ($B$) states form a near-degenerate doublet. 
In the trigonometric case these doublet states resemble free rotor states above the barrier.
In the hyperbolic case, they form a tunneling doublet below the barrier.

In the field-free limit, $\beta=0$, the analytic energies given in Table~\ref{tab:ana_energies} simplify to
\begin{equation}
\label{eq:beta0}
	E_{t,\kappa}^{(\Gamma)} = - E_{h,\kappa}^{(\Gamma)} \in \left\{\left(\frac{\nu}{2}\right)^2, \, 0 \le |\nu| \le \kappa-1\right\}
\end{equation}
with even $\nu$ for $A_1$ and $A_2$ states (for odd $\kappa$) or odd $\nu$ for $B_1$ and $B_2$ states (for even $\kappa$).
Note that  a $\nu=0$ state exists only for $A_1$.
These eigenvalues can also be found by directly inserting $\beta=0$ in all four $T_{\kappa}^{(\Gamma)}$ matrices, Eqs.~(\ref{eq:T_t_A1})-(\ref{eq:T_t_A2}).
Then all subdiagonal elements vanish, thereby rendering these matrices exactly solvable with the above eigenvalues.
Alternatively, one can also arrive at the same solutions by setting $\beta=0$ in Eq.~(\ref{eq:T_t}) or (\ref{eq:T_h}), in which case they become Chebychev (type I) equations.
  
For the pendular system in the field-free ($\beta=0$) limit, these results can be simply understood as the energy levels of a free rotor, but with the quantum number $\nu$ divided by two in order to account for the periodicity which is here 4$\pi$ instead of the usual 2$\pi$.

For the hyperbolic counterpart, however, it is not possible to reach the $\beta=0$ limit continuously.
Instead, we consider the limit of $\beta \approx 0$, where the Razavy potential takes the form of a double Morse potential \cite{Ulyanov1992v}
\begin{equation}
\label{eq:Raz_DM}
	\lim_{\beta\rightarrow 0}V_h(x)=\frac{\kappa^2}{4}\left(1-e^{-\left(x+\ln\left(\frac{\kappa}{|\beta|}\right)\right)}\right)^2+ \frac{\kappa^2}{4}\left(1-e^{\left(x-\ln\left(\frac{\kappa}{|\beta|}\right)\right)}\right)^2-\frac{\kappa^2}{2}
\end{equation}
Here the distance $d=2\ln (\kappa/|\beta|)$ of the two wells increases with decreasing $|\beta|$, but the dissociation energy depends only on $\kappa$, see also Section~\ref{sec:ham_h}.
For each of the two Morse oscillators alone ($d\rightarrow\infty$), the energy levels are exactly as given above in Eq.~(\ref{eq:beta0}).
All these states are bound states except for $\nu=0$ ($A_1$), the energy of which coincides with the dissociation threshold. 
By decreasing the distance $d$ (increasing $|\beta|$) between the two wells of the double Morse oscillator, Eq.~(\ref{eq:Raz_DM}), the energy levels will increasingly perturb one another and near-degenerate tunneling doublets will eventually form. 

\subsubsection{Near-degenerate doublets}
For small $|\beta|$-values, the degeneracies found for $\beta=0$ are lifted and instead near-degenerate doublets are formed, see again  Figures~\ref{fig:k5_b075} and \ref{fig:k6_b075}.
These doublets are found near the free rotor limit of the trigonometric system or as tunneling doublets in the hyperbolic system. 
The corresponding splittings can be derived from Table~\ref{tab:ana_energies}.
Because they apply equally to the two classes of systems, we will drop the {\it t} and {\it h} subscripts on the energies. 
For the simplest example ($\kappa=2$), we find by using Eq.~(\ref{eq:E_k1_A1})
\begin{equation}
  \label{eq:split_k2}
	\left|(E_2^{(B_1)})_0-(E_2^{(B_2)})_0\right|=2|\beta|
\end{equation}

Similary, for $\kappa=3$, the splitting between the lowest two states ($A_1$ and $A_2$) as obtained from (\ref{eq:E_k3_A2}) and (\ref{eq:E_k3_A1}) is
\begin{equation}
  \label{eq:split_k3}
	\left|(E_3^{(A_2)})_0-(E_3^{(A_1)})_0\right|=4\beta^2 +\mathcal{O}(\beta^4)
\end{equation}
where the third power, as well as all other odd powers, of $\beta$ vanish identically.
Note that the third analytic state, $\nu=0$ ($A_1$) in Eq.~(\ref{eq:E_k3_A1}), already lies above the barrier.
For $\kappa=4$, the four analytic states comprise two tunneling doublets with energy splittings
\begin{eqnarray}
  \label{eq:split_k4}
	\left|(E_4^{(B_1)})_0-(E_4^{(B_2)})_0\right|&=&3|\beta|^3  +\mathcal{O}(\beta^5) \nonumber \\
	\left|(E_4^{(B_1)})_1-(E_4^{(B_2)})_1\right|&=&4|\beta| + 3|\beta|^3 +\mathcal{O}(\beta^5)
\end{eqnarray}
where the splitting of the upper doublet is much larger than the lower one for $\beta\rightarrow 0$.

\begin{figure}
\includegraphics[height=12cm]{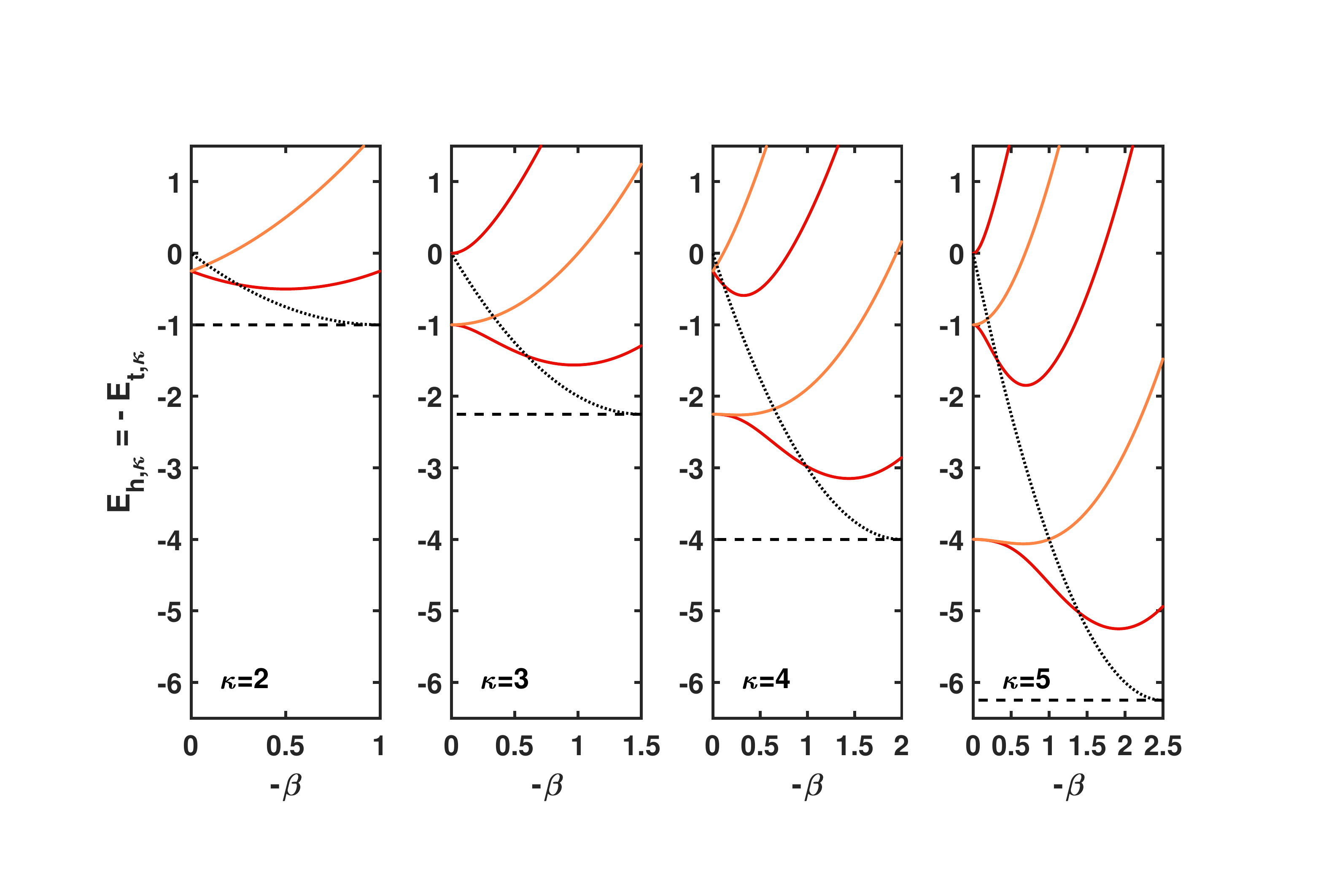}
\caption{Analytic energy levels of the trigonometric ($-E_{t,\kappa}$) and hyperbolic ($E_{h,\kappa}=-E_{t,\kappa}$) system for small values of $|\beta|$ for $2 \le \kappa \le 5$. In the limit of $\beta\rightarrow 0$ there are $\kappa/2$ doublets for even $\kappa$ or $(\kappa-1)/2$ doublets for odd $\kappa$. The dashed lines indicate the minima of the Razavy potential. The dotted curves show the maximum of the Razavy potential, or the negative of the maxima of the pendular potential.}
\label{fig:splitting}
\end{figure}

For higher $\kappa$ this pattern for the analytic states in the limit of small $\beta$ continues. 
For even $\kappa$, there are always $\kappa/2$ doublets.
For odd $\kappa$, there are only $(\kappa-1)/2$ doublets  whereas the highest single $A_1$ state lies always above the barrier, see Figure~\ref{fig:splitting}.
As already mentioned, the splittings increase with the energies of the doublets.
With increasing $|\beta|$, the splittings grow larger, and the doublets become single states.
Typically, this behavior is found where the energy curves cross the black dotted curves also shown in Figure~\ref{fig:splitting}.
For the pendular system this means that the energies fall below the maxima of the potential, where the transition from a (nearly) free rotor to a librator (hindered rotor) takes place.
For the Razavy system this corresponds to the energies exceeding the potential barrier of the double well, i.e., the transition from tunneling to a single oscillator.

\section{numerical solutions}
\label{sec:num}
Up to this point we discussed the analytic eigenproperties of the finite, $N_{\kappa}^{(\Gamma)}$-dimensional blocks of the matrices given in Eqs.~(\ref{eq:T_t_A1}) - (\ref{eq:T_t_A2}). 
However, these solutions were restricted to the case of odd (or even) integer $\kappa$ for periodicity pertaining to the $A$ (or $B$) symmetry,  because only in those cases the infinite-dimensional matrices given in Eqs.~(\ref{eq:T_sup})-(\ref{eq:T_sub}) could be broken into two blocks each, due to the presence of a single zero entry in the respective subdiagonals. 
In this section, we go beyond the C-QES (and AIS) solution spaces and consider the complete spectra of the pendular and the Razavy systems.

\subsection{Numerical diagonalization of truncated tridiagonal matrices}
The tridiagonal matrices Eqs.~(\ref{eq:T_sup})-(\ref{eq:T_sub}) can also be used to obtain the eigenvalues and eigenfunctions of the trigonometric system numerically. The accuracy of the eigenproperties depends on the dimension of the matrices used for the numerical diagonalization, i.e., on their truncation (typically at a dimension of a few hundred, depending on the magnitude of $\beta$). Table~\ref{tab:num_pendulum} provides a list of the numerical pendular eigenenergies. 
The resulting eigenvectors can be used to construct the corresponding eigenfunctions, again as products of von Mises functions, seed functions, and (even ordered) polynomials in $\cos(\theta/2)$, cf. Eq.~(\ref{eq:eig_t_AB}), \begin{eqnarray}
\label{eq:eig_ext_AB}
\left(\psi_{t,\kappa}^{(A_1)}\right)_n \propto                      
e^{\beta\cos\theta} \left(
\sum_{\ell=0}^{(\kappa-1)/2} \left( v_{\kappa}^{(A_1)}\right)_{n,\ell} \cos^{2\ell}  \frac{\theta}{2} +
\sum_{\ell=(\kappa+1)/2}^\infty \left( v_{\kappa}^{(A)}\right)_{n,\ell} \cos^{2\ell}  \frac{\theta}{2} 
\right) \nonumber \\
\left(\psi_{t,\kappa}^{(B_1)}\right)_n \propto                      
e^{\beta\cos\theta} 
\cos\frac{\theta}{2}  \left(
\sum_{\ell=0}^{(\kappa-2)/2} \left( v_{\kappa}^{(B_1)}\right)_{n,\ell} \cos^{2\ell}\frac{\theta}{2} +
\sum_{\ell=\kappa/2}^\infty \left( v_{\kappa}^{(B)}\right)_{n,\ell} \cos^{2\ell}\frac{\theta}{2} 
\right) \nonumber \\
\left(\psi_{t,\kappa}^{(B_2)}\right)_n \propto  
e^{\beta\cos\theta} 
\sin\frac{\theta}{2}  \left(
\sum_{\ell=0}^{(\kappa-2)/2} \left( v_{\kappa}^{(B_2)}\right)_{n,\ell} \cos^{2\ell}  \frac{\theta}{2} + 
\sum_{\ell=\kappa/2}^\infty \left( v_{\kappa}^{(B)}\right)_{n,\ell} \cos^{2\ell}  \frac{\theta}{2} 
\right) \nonumber \\
\left(\psi_{t,\kappa}^{(A_2)}\right)_n \propto  
e^{\beta\cos\theta} 
\sin\theta  \left(
\sum_{\ell=0}^{(\kappa-3)/2} \left( v_{\kappa}^{(A_2)}\right)_{n,\ell} \cos^{2\ell}\frac{\theta}{2} +
\sum_{\ell=(\kappa-1)/2}^\infty \left( v_{\kappa}^{(A)}\right)_{n,\ell} \cos^{2\ell}\frac{\theta}{2}
\right) \,.
\end{eqnarray}
It is known from the literature  on the Whittaker-Hill equation \cite{Winkler1958,Magnus2004w} that the above expansions can be split: 
While the first $N_\kappa^{(\Gamma)}$ columns are different for each of the four irreducible representations, the remaining columns are the same for the two classes of $2\pi$-periodic solutions ($A_1$ and $A_2$) and also for the $2\pi$-antiperiodic solutions ($B_1$ and $B_2$). 
Hence, there is no need for subscripts 1 or 2 on the irreducible representations  denoting the $v$ matrices in the second terms of the above equations.

However, for the hyperbolic system an \textit{ansatz} equivalent to Eq.~(\ref{eq:eig_ext_AB}), but with the trigonometric functions replaced by their hyperbolic counterparts, results in non-normalizable wavefunctions.
Unlike the finite-dimensional case discussed in Section~\ref{sec:solve}, the infinite sums lead to a strong divergence for $x\rightarrow \pm \infty$, because the cosh functions outweigh the hyperbolic von Mises function (even for $\beta<0$). 
We note that this problem is connected with the anti-isospectral transform given by Eq.~(\ref{eq:ais}).
While the solutions to the trigonometric problem are square-integrable for $-2 \pi \le \theta \le 2 \pi$, this mapping renders the solutions of the hyperbolic problem square-integrable on the $-2 \mathrm{i} \pi \le x \le 2 \mathrm{i} \pi$ interval, rather than  $-\infty \le x \le \infty$, as required for the Razavy potential.
Hence, the approach outlined in Section~\ref{sec:solve_h}, based on the tridiagonal matrices~(\ref{eq:T_sup})-(\ref{eq:T_sub}), is not suitable for generating states beyond the range of the analytic solutions of the hyperbolic system.
Instead, we use the Fourier Grid Hamiltonian (FGH) approach \cite{Meyer:70a,Marston:89a} implemented in the \textsl{qm\_bound} program of the \textsc{WavePacket} software package \cite{BSchmidt:75}.
Within the energy ranges considered here, well-converged energies are obtained using 1024 equally spaced grid points.

These numerical techniques allow us to calculate two types of energy levels which could not be obtained with the analytic methods developed in Section~\ref{sec:solve}.
Firstly, for the pendular system, these are the $B_1$ and $B_2$ symmetry states for odd values of $\kappa$ as well as the $A_1$ and $A_2$ symmetry states for even values of $\kappa$.  
As can be seen in Table~\ref{tab:num_pendulum}, these values barely differ from their analytic counterparts. 
However, this is only due to the rather large value of $|\beta|=5$ chosen here; for other values, see below.
Secondly, Tables~\ref{tab:num_pendulum} and \ref{tab:num_razavy} also contain energy levels with $n\ge N_\kappa^{(\Gamma)}$, i.e., beyond the energetically highest analytic states. 
A more thorough discussion of this part of the spectra as well as the spectra obtained for non-integer values of $\kappa$ is given below.

\subsection{Anti-isospectrality}

Figures~\ref{fig:e_b500} and \ref{fig:e_b075} show, respectively, the energy levels of both the pendular and the Razavy systems as continuous function of $\eta$ (or $\kappa$) for large and small values of $|\beta|$.
In both cases the lower dashed lines show the global minimum of the pendular potential.
The upper dashed lines/curves show the negative of the minimum/minima of the Razavy potential (which coincide with the local minimum of the pendulum in case of $|\beta|>\kappa/2$ only).
These boundaries define the interval of quasi-exact solvability, $\left[V_t(\theta_{min,g}) , -V_h(x_{min})\right]$.
As indicated by the black circles in Figures \ref{fig:e_b500} and \ref{fig:e_b075}, all analytic eigenvalues obtained by the methods of Section~\ref{sec:solve} are restricted to this interval.
As required by the anti-isospectrality, Eq.~(\ref{eq:ais_E}), these circles are found at the crossings of the energy curves for the pendular system (green and blue) with the negative energy curves of the Razavy system (red and orange).
These crossings are located at odd integer $\kappa$ for the $A$ states of the pendulum ($A_1\leftrightarrow A'$,  $A_2\leftrightarrow A''$) or even values of $\kappa$ for the $B$ states of the pendulum ($B_1\leftrightarrow A'$, $B_2\leftrightarrow A''$), see Figures~\ref{fig:e_b500} and \ref{fig:e_b075}.
Note that between these values of $\kappa$ the anti-isospectrality does not hold. This includes the case of the other crossings in Figures~\ref{fig:e_b500} and \ref{fig:e_b075} that are only accidentally close to those for even values of $\kappa$ for $A$ states and odd values of $\kappa$ for $B$ states.

\subsection{Genuine and avoided crossings}

Finally, we discuss the spectral structures of each of the potentials separately.
The Schr\"{o}dinger Eq.~(\ref{eq:se_h}) for the Razavy system is a non-periodic Sturm-Liouville equation.
Hence the  energy levels are non-degenerate and can be ordered as
\begin{equation} \label{spec_h}
\left(E_{h,\kappa}^{(A')}\right)_0 < 
\left(E_{h,\kappa}^{(A'')}\right)_1 < 
\left(E_{h,\kappa}^{(A')}\right)_2 < 
\left(E_{h,\kappa}^{(A'')}\right)_3 < 
\left(E_{h,\kappa}^{(A')}\right)_4 < \ldots  
\end{equation} 
where -- from now on -- the numbering of the energy levels $n=0,1,2,\ldots$ is irrespective of the irreducible representations. 
This strictly monotonic growth scheme also applies to the near degenerate energy levels in the upper part of Figure~\ref{fig:e_b075}, see discussion at the end of Section~\ref{sec:limits}. 
There, the tuneling doublets lie below the barrier of the Razavy potential and are separated from other energy levels above the maximum of the potential which are all single states.

In contrast, the Schr\"{o}dinger Eq.~(\ref{eq:se_t}) for the trigonometric system is a periodic Sturm-Liouville equation.
Hence, the oscillation theorem can be applied.
This theorem classifies the eigenvalues of such an equation with respect to the periodic and anti-periodic boundary conditions of their corresponding eigenfunctions \cite{Teschl2012g,Gerlach2015u,Magnus2004w}.
In particular, it states that the spectrum of the planar quantum pendulum is purely discrete and there is a sequence of real eigenvalues 
\begin{equation} \label{spec_t}
\left(E_{t,\kappa}^{(A)}\right)_0 < 
\left(E_{t,\kappa}^{(B)}\right)_1 \leq 
\left(E_{t,\kappa}^{(B)}\right)_2 < 
\left(E_{t,\kappa}^{(A)}\right)_3 \leq 
\left(E_{t,\kappa}^{(A)}\right)_4 < 
\left(E_{t,\kappa}^{(B)}\right)_5 \leq 
\left(E_{t,\kappa}^{(B)}\right)_6 < \ldots 
\end{equation} 
where, again, the numbering of the energy levels is irrespective of the irreducible representations. 
No distinction between $A_1$ and $A_2$ ($B_1$ and $B_2$) is given here because there is no unique pattern.
The first $2\kappa$ levels are within the interval of quasi-exact solvability and are non-degenerate (single states); their ordering pattern is ($A_1$, $B_1$, $B_2$, $A_2, \ldots$).
Above this interval, the ordering pattern changes, which is connected with the genuine and avoided crossings in the upper part of the spectra shown in Figures~\ref{fig:e_b500} and \ref{fig:e_b075}.
Genuine crossings are found for odd $\kappa$ for all $A$ states beyond the $\kappa$ analytic, single states.
Similarly, for $B$ states these degeneracies appear only for even $\kappa$.
This is the expected behavior due to the coexistence theorem \cite{Winkler1958,Magnus2004w, Hemery2010a} for the Whittaker-Hill Eq.~(\ref{eq:se_t}), which states that there can be pairs of linearly independent coexisting  solutions, of period $2\pi$ for the same value of $E_{t,\kappa}^{(A_1)}=E_{t,\kappa}^{(A_2)}$ if and only if $\kappa$ is an odd integer.
Similarly, pairs of $2\pi$-antiperiodic solutions can coexist for eigenenergy $E_{t,\kappa}^{(B_1)}=E_{t,\kappa}^{(B_2)}$ if and only if $\kappa$ is an even integer. 
Moreover, as a consequence of the coexistence theorem \cite{Magnus2004w, Hemery2010a}, one can predict that these degeneracies will appear for all states above the interval of QES.

There are also avoided crossings above the interval of QES in the upper part in Figs.~\ref{fig:e_b500} and \ref{fig:e_b075}.
For $B$ states they are found for odd $\kappa$, whereas for $A$ states they occur for even $\kappa$.
In accordance with the Wigner-von Neumann theorem (non-crossing rule), these avoided crossings involve states pertaining to the same irreducible representations.
Note that some of the gaps cannot be discerned on the scale of the figures.
Nevertheless, considering Eq.~(\ref{spec_t}), it is apparent that there is a degenerate pair of $A$ states inside every gap between $(E_{t,\kappa}^{(B)})_n$ and $(E_{t,\kappa}^{(B)})_{n+3}$ states for even $\kappa$ and vice versa for odd $\kappa$.
As a consequence of the genuine and avoided crossings the pattern of energy levels beyond the QES interval is ($A_1$, $B_1$=$B_2$, $A_1$, $A_2$, $B_1$=$B_2$, $A_2,\ldots$) or  ($B_2$, $A_1$=$A_2$, $B_2$, $B_1$, $A_1$=$A_2$, $B_1,\ldots$).

\section{Conclusions and Prospects}
\label{sec:con} 

We showed that the planar pendulum and the Razavy system possess symmetries isomorphic with those of the point groups $C_{2v}$ and $C_{i}$, whereby the irreducible representations $A_1$, $B_1$  and $A_2$, $B_2$  of  $C_{2v}$ correlate with the irreducible representation $A'$  and $A''$ of $C_i$, respectively. We found that the  analytic solutions reported in Refs. \cite{Schmidt2014b} and  \cite{Razavy1980m} for the lowest states of the two systems indeed exhibit these symmetries. Furthermore, we found a total of forty analytic solutions for the planar pendulum and determined that even and $2\pi$-periodic solutions correspond to the $A_1$ symmetry, odd and $2\pi$-periodic solutions to $A_2$, even and $2\pi$-antiperiodic solutions to $B_1$, and odd and $2\pi$ anti-periodic solutions to $B_2$ symmetry. For the Razavy system we found that the solutions are non-periodic, of even or odd paritiy for the $A'$ or $A''$ symmetry, respectively. For the dimensionless interaction parameters $\eta$ and $\zeta$ such that $|\eta| > 2\zeta$, the pendular potential is a single well whereas the Razavy potential is a double well, provided that, in addition, $\eta<0$. Conversely, for $|\eta| < 2\zeta$, the pendular potential is a double well and the Razavy potential a single well, provided $\eta<0$. 

In Ref.~\cite{Schmidt2014b}, we showed that the topology of the intersections (genuine or avoided) of the planar pendulum's eigenenergy surfaces, spanned by $\eta$ and $\zeta$, can be characterized by a single integer index $\kappa$ (the topological index) and that the values of $\kappa$ correspond to the sets of conditions imposed on $\eta$ and $\zeta$ under which analytic solutions of the planar quantum pendulum problem obtain. The parabolic surfaces running through the loci of the intersections for a given $\kappa$ can be termed {\it parabolae of conditional quasi-exact solvability}. 
In the present work we were able to trace the origin of the parabolae of quasi-exact solvability to the structure of the tridiagonal matrices representing the symmetry-adapted pendular Hamiltonian: If and only if $\kappa$ is an odd/even positive integer can the tridiagonal matrices, each of which corresponds to one of the problem's four irreducible representations, be broken into finite-dimensional matrices and infinite-dimensional remainders, whereby the finite-dimensional matrices can be diagonalized, at least in principal, analytically, with solutions that are periodic/antiperiodic in $2\pi$. The dimensions of the finite block matrices add up to the topological index $\kappa$, which, therefore, equals the number of analytic solutions. Although we can find, in principle, infinitely many analytic solutions, we cannot find all solutions analytically. In particular, the solutions that remain out of reach are those that correspond to either $\eta$ or $\zeta$ equal to zero (i.e., no analytic solutions to the Mathieu equation obtain). For non-integer $\kappa$, the tridiagonal matrices are infinite and, therefore, not amenable to analytic diagonalization.

We have shown that, despite the rather different symmetries and irreducible representations, the pendular and Razavy Hamiltonians can be represented by the {\it same} four tridiagonal matrices, cf. also \cite{lopez1993}. Hence the exactly solvable parts of their spectra are the same (up to a sign), which renders the pendular and Razavy systems anti-isospectral (AIS).  The iso-spectrality occurs for single states only (i.e., not for doublets). Moreover, at a given $\kappa$, the anti-isospectrality occurs for single states only (i.e., not for doublets), like C-QES  holds solely for integer values of $\kappa$, and only occurs for the lowest eigenvalues of the pendular and Razavy Hamiltonians, with the order of the eigenvalues reversed for the latter.  For all other states, the pendular and Razavy spectra become in fact qualitatively different, as higher pendular states appear as doublets and higher Razavy doublets appear as single states. 

The study of the two-dimensional (2D) planar pendulum proved its worth in providing inspiration for solving the full-fledged three-dimensional (3D) pendulum eigenproblem, cf. Refs. \cite{LemeshkoPRA2011,LemeshkoNJP2011,Schmidt2014b, friedrich2015}. In particular, the lowest 2D solutions could be used as {\it Ansatz} for the superpotentials \cite{Cooper1995f,Robnik1997m} on which the search for analytic solutions via supersymmetric quantum mechanics relies. Equipped with many more analytic solutions in 2D, this search will continue. 

Last but not least, the analytic solvability of the time-dependent pendular eigenproblem \cite{Ortigoso1999, CaiFri2001,Leibscher2004,Owschimikow2009,Owschimikow2011} -- both in 2D and 3D -- will be investigated in pursuit of dynamical models of the interactions of molecules with electric, magnetic and optical fields. 

\begin{acknowledgments}
Support by the Deutsche Forschungsgemeinschaft (DFG) through grants  SCHM 1202/3-1 and FR 3319/3-1 is gratefully acknowledged.
\end{acknowledgments} 

\newpage

\bibliography{PlanRaz} 

\clearpage

\begin{turnpage}
\begin{table}
\begin{ruledtabular}
\begin{tabular}{l l  l l}
		\multicolumn{2}{c}{Pendular trigonometric potential} 
	&
		\multicolumn{2}{c}{Razavy hyperbolic potential}
	\\ \noalign{\smallskip}	\hline \noalign{\smallskip} 
			Problems \& Applications
		& 	Reference 
		&  	Problems \& Applications 
		& 	Reference
	\\\noalign{\smallskip} \hline \noalign{\smallskip}
			Internal rotation, molecular torsion 
		& 	\cite{Herschbach_1957,Herschbach_1959,ramakrishna2007,Parker2011s,Bitencourt2008a,Roncaratti2010l} 
		& 	Uniaxial paramagnets or single-molecule magnets 
		& 	\cite{yunbo1998, Ulyanov1984v,ulyanov1997} 
	\\
			Molecular alignment/orientation 
		& 	\cite{Friedrich_1991,Leibscher-Schmidt_2009} 
		& 	Quasi-exactly solvable double well potential 
		&	\cite{Hemery2010a,Finkel1999f,Turbiner1988a} 
	\\    
			Molecules in combined fields 
		& 	\cite{Schmidt2014b} 
		&	Quantum field theory 
		& 	\cite{Khare1998a,Habib1998s,Konwent1998h} 
	\\
			Band structure of condensed matter systems 
		& 	\cite{Khare2004a,Hemery2010a,Finkel1999f} 
		& 	PT-symmetry
		& 	\cite{bagchi2001}  
	\\  
			Nonlinear dynamics 
		& 	\cite{Condat1983c, Campbell1986d}
		&	Nonlinear coherent structures 
		& 	\cite{Khare1997a,Behera1981s} 
	\\
		Solitons
		&	\cite{burt1978,Kar2000s} 
		& 	Quantum theory of instantons
		& 	\cite{AIzquierdo2014a,Kar2000s,Razavy2003m}  
	\\    
			Josephson Junction Rhombus 
		& 	\cite{ulrich2015} 
		& 	Model of a proton in a	hydrogen bond 
		& 	\cite{Finkel1999f,konwent1986,Konwent1998h}  
	\\ 
			Optical lattice 
		& 	\cite{graham1992} 
		&  
		&   
\end{tabular}
\end{ruledtabular}
\caption{Examples of problems and applications in chemistry and physics where the pendular trigonometric potential, Eq. (\ref{eq:pot_t}), and the Razavy hyperbolic potential, Eq. (\ref{eq:pot_h}), make a prominent appearance.}
\label{tab:examples}
\end{table}
\end{turnpage}

\begin{table}
\begin{ruledtabular}
\begin{tabular}{c c c c c | c c c}
        \multicolumn{5}{c}{Pendulum}
    &
        \multicolumn{3}{c}{Razavy system}
    \\    \hline
        $\Gamma_{t}$ & $E$ & $R(2\pi)$ & $P(0)$ & $P(\pi)$
    &    $\Gamma_{h}$ & $E$ & $P$
    \\ \hline
            $A_1$ &     1 & 1 & 1 & 1
        &
            \multirow{2}{*}{$A'$ } & \multirow{2}{*}{1} & \multirow{2}{*}{1}\\
            $B_1$ &     1 & -1 & 1 & -1  &  &   &    \\
            $B_2$ &     1 & -1 & -1 & 1
        &
            \multirow{2}{*}{$A''$ } & \multirow{2}{*}{1} & \multirow{2}{*}{-1}\\
            $A_2$ &     1 & 1 & -1 & -1 &   &   &
    \\
\end{tabular}
\end{ruledtabular}
\caption{Character tables for the irreducible representations of the planar pendulum (trigonometric, $C_{2v}$) and Razavy (hyperbolic, $C_i$) systems. The symmetry operations are defined by Eqs.~(\ref{eq:map_C2v}) and (\ref{eq:map_Ci}).}
\label{tab:irreps}
\end{table}

\begin{table}
\begin{ruledtabular}
\begin{tabular}{cccc|cccc}
$\kappa$ &  $\Gamma_t$ & $E_{t,\kappa}^{(\Gamma_t)}$ & $\psi_{t,\kappa}^{(\Gamma_t)}(\theta)\propto$ & $\kappa$ & $\Gamma_h$ & $E_{h,\kappa}^{(\Gamma_h)}$ & $\psi_{h,\kappa}^{(\Gamma_h)}(x)\propto$ \\
\hline
1 & $A_1$ &  $-\beta^{2}$  &  $\mathrm{e}^{\beta \cos\theta}$ & \multirow{2}{*}{1} & \multirow{2}{*}{$A'$} & \multirow{2}{*}{$\beta^{2}$}  & \multirow{2}{*}{$\mathrm{e}^{\beta \cosh x}$} \\
2  &  $B_1$ & $-\beta^{2}-\beta+\frac{1}{4}$ & $\cos\frac{\theta}{2}\mathrm{e}^{\beta\cos\theta}$ & & & & \\
2 &   $B_2$ & $-\beta^{2}+\beta+\frac{1}{4}$ & $\sin\frac{\theta}{2}\mathrm{e}^{\beta\cos\theta}$ & \multirow{2}{*}{2} & \multirow{2}{*}{$A''$} & \multirow{2}{*}{$\beta^{2}-\beta-\frac{1}{4}$} & \multirow{2}{*}{$\sinh\frac{x}{2}\mathrm{e}^{\beta \cosh x}$}\\
3 &   $A_2$ & $-\beta^{2}+1$ & $\sin\theta\mathrm{e}^{\beta\cos\theta}$ &  &  &   &
\end{tabular}
\end{ruledtabular}
\caption{Analytic eigenenergies and wavefunctions of the four lowest states of the pendulum (trigonometric) and Razavy (hyperbolic) systems as identified in Refs.~\cite{Schmidt2014b} and \cite{Razavy1980m}, respectively. We note that in Razavy's original work \cite{Razavy1980m} an integer $n$ is used such that $\kappa=n+1$ and interaction parameter $\xi$ such that $\zeta =\frac{\xi^2}{16}\equiv  \beta^2$.}

\label{tab:seed}
\end{table}

\begin{table}
\begin{ruledtabular}
\begin{tabular}{c c c c l}
$\kappa$           & $\Gamma_t$ & $\Gamma_h$ & $n$ & $E_{t,\kappa}^{(\Gamma)}=-E_{h,\kappa}^{(\Gamma)}$        \\
\hline
                1  & $\underline{A_1}$ & $\underline{A'}$ & 0 & $-\beta^2$ \\  \hline
\multirow{2}{*}{2} & $\underline{B_1}$ & $A'$  & 0 & $-\beta^2 - \beta + \frac{1}{4}$ \\
                   & $\underline{B_2}$ & $\underline{A''}$ & 0 & $-\beta^2 + \beta + \frac{1}{4}$ \\ \hline
\multirow{3}{*}{3} & $A_1$ & $A'$  & 0 & $-\beta^2 - \frac{1}{2}\sqrt{16\beta^2 + 1} + \frac{1}{2} $  \\
                   & $A_1$ & $A'$  & 1 & $-\beta^2 + \frac{1}{2}\sqrt{16\beta^2 + 1} + \frac{1}{2}$  \\
                                     & $\underline{A_2}$ & $A''$ & 0 & $-\beta^2 + 1$ \\\hline
\multirow{4}{*}{4} & $B_1$ & $A'$  & 0 & $-\beta^2 - \beta - \sqrt{4\beta^2 + 2\beta + 1} + \frac{5}{4}$ \\
                   & $B_1$ & $A'$  & 1 & $-\beta^2 - \beta + \sqrt{4\beta^2 + 2\beta + 1}  + \frac{5}{4}$ \\
                   & $B_2$ & $A''$ & 0 & $-\beta^2 + \beta - \sqrt{4\beta^2 - 2\beta + 1}  + \frac{5}{4}$ \\
                   & $B_2$ & $A''$ & 1 & $-\beta^2 + \beta + \sqrt{4\beta^2 - 2\beta + 1} + \frac{5}{4}$ \\  \hline
\multirow{5}{*}{5} & $A_1$ & $A'$  & 0 & $-\beta^2 - \frac{1}{3a}(a^2+ 48 \beta^2 + 13) + \frac{5}{3}  $ \\
                   & $A_1$ & $A'$  & 1 & $-\beta^2 + \frac{1}{6a}(a^2+ 48 \beta^2 + 13) - \frac{i\sqrt{3}}{6a}(-a^2+ 48 \beta^2 + 13) +\frac{5}{3} $ \\
                   & $A_1$ & $A'$  & 2 & $-\beta^2 + \frac{1}{6a}(a^2+ 48 \beta^2 + 13) + \frac{i\sqrt{3}}{6a}(-a^2+ 48 \beta^2 + 13) +\frac{5}{3} $ \\
                                     & $A_2$ & $A''$ & 0 & $-\beta^2 - \frac{1}{2}\sqrt{16\beta^2 + 9}  + \frac{5}{2}$ \\
                   & $A_2$ & $A''$ & 1 & $-\beta^2 + \frac{1}{2}\sqrt{16\beta^2 + 9}  + \frac{5}{2}$ \\ \hline
\multirow{6}{*}{6} & $B_1$ & $A'$  & 0 & $-\beta^2 - \beta -\frac{1}{3b_+}(b_+^2 + 48\beta^2+24\beta + 28) + \frac{35}{12}$ \\
                   & $B_1$ & $A'$  & 1 & $-\beta^2 - \beta +\frac{1}{6b_+}(b_+^2 + 48\beta^2+24\beta+28)-\frac{i\sqrt{3}}{6b_+}(-b_+^2 + 48\beta^2+24\beta+28) + \frac{35}{12}$ \\
                   & $B_1$ & $A'$  & 2 & $-\beta^2 - \beta +\frac{1}{6b_+}(b_+^2 + 48\beta^2+24\beta+28)+\frac{i\sqrt{3}}{6b_+}(-b_+^2 + 48\beta^2+24\beta+28) + \frac{35}{12}$ \\
                                     & $B_2$ & $A''$ & 0 & $-\beta^2 + \beta -\frac{1}{3b_-}(b_-^2 + 48\beta^2-24\beta + 28) + \frac{35}{12}$ \\
                                     & $B_2$ & $A''$ & 1 & $-\beta^2 + \beta +\frac{1}{6b_-}(b_-^2 + 48\beta^2-24\beta+28)-\frac{i\sqrt{3}}{6b_-}(-b_-^2 + 48\beta^2-24\beta+28) + \frac{35}{12}$ \\
                   & $B_2$ & $A''$ & 2 & $-\beta^2 + \beta +\frac{1}{6b_-}(b_-^2 + 48\beta^2-24\beta+28)+\frac{i\sqrt{3}}{6b_-}(-b_-^2 + 48\beta^2-24\beta+28) + \frac{35}{12}$ \\\hline
\multirow{3}{*}{7} & $A_2$ & $A''$ & 0 & $-\beta^2 - \frac{1}{3d}(d^2+48\beta^2+49) + \frac{14}{3}$\\
                   & $A_2$ & $A''$ & 1 & $-\beta^2 + \frac{1}{6d}(d^2+48\beta^2+49) -\frac{i\sqrt{3}}{6d}(-d^2+48\beta^2+49) + \frac{14}{3}$ \\
                   & $A_2$ & $A''$ & 2 & $-\beta^2 + \frac{1}{6d}(d^2+48\beta^2+49) +\frac{i\sqrt{3}}{6d}(-d^2+48\beta^2+49) + \frac{14}{3}$

\end{tabular}
\end{ruledtabular}
\caption{\scriptsize{Analytical energies of the planar pendulum (trigonometric, $E_t$) and the Razavy system (hyperbolic, $-E_h$). Here $\Gamma_t$ and $\Gamma_h$ stand for the irreducible representations of the $C_{2v}$ and $C_i$ point groups and  $a=\sqrt[3]{ 6\sqrt{3(- 1024\beta^6 - 64\beta^4 - 412\beta^2 - 9)} + 288\beta^2 - 35}$, $b_{\pm}=\sqrt[3]{288\beta^2 \pm 36\beta-80+12 c_{\mp}}$, $c_{\pm}=\sqrt{3(- 256\beta^6 \pm 384\beta^5 - 448\beta^4 \pm 432\beta^3 - 477\beta^2 \pm 144\beta - 36)}$ , $d=\sqrt[3]{12\sqrt{3\left(-256\beta^6-592\beta^4-991\beta^2-225\right)}+288\beta^2-143} $.  In some cases $a , b_{\pm}$ are complex numbers but nonetheless imaginary parts of all eigenenergies vanish. Irreducible representations of seed functions are underlined.}}
\label{tab:ana_energies}
\end{table} 

\begin{turnpage}
\begin{table}
\begin{ruledtabular}
\begin{tabular}{cccccccccccc}
\multicolumn{2}{c}{$\kappa=1$} & \multicolumn{2}{c}{$\kappa=2$} & \multicolumn{2}{c}{$\kappa=3$} & \multicolumn{2}{c}{$\kappa=4$} & \multicolumn{2}{c}{$\kappa=5$} & \multicolumn{2}{c}{$\kappa=6$}\\
\hline
A & B &  A & B & A & B &  A & B & A & B & A & B \\
\hline
\textbf{-25} & \textit{-25.0000} & -29.7500 & \textit{\textbf{-29.75}} & \textbf{-34.5125} & \textit{-34.5125} & -39.2857 & \textit{\textbf{-39.2857}} &  \textbf{-44.0681} & \textit{-44.0681} & -48.8587 & \textit{\textbf{-48.8587}} \\
\multirow{2}{*}{\underline{-15.5485}} &-15.5601 & \textit{-19.7500} & \textbf{-19.75} & \textbf{\textit{-24}} & -24.0000 &\textit{-28.2894} & \textbf{-28.2894} & \textbf{\textit{-32.6119}} & -32.6119 & \textit{-36.9628}& \textbf{-36.9628} \\
& -15.5369 & -10.8997 & \multirow{2}{*}{\underline{-10.8565}} & \textbf{-14.4875} & \textit{-14.4875}  & -18.2143 & \textbf{\textit{-18.2143}} & \textbf{-22.0150} & \textit{-22.0150} & -25.8760 & \textit{\textbf{-25.8760}} \\
~\multirow{2}{*}{\underline{-7.3631}}  & ~\textit{-7.5512} & -10.8118 &  & ~\multirow{2}{*}{\underline{-6.1992}} & ~-6.3212 & ~\textit{-9.2105} & ~\textbf{-9.2106}& \textbf{\textit{-12.3881}} & -12.3881 & \textit{-15.6840} & \textbf{-15.6840} \\
 & ~\textit{-7.1487} & ~\textit{-3.8735} & ~\multirow{2}{*}{\underline{-3.4452}} &  & ~-6.0650 & ~-1.8613 & ~\multirow{2}{*}{\underline{-1.5925}} & ~\textbf{-3.9169} & ~\textit{-3.9161} & ~-6.5154 & ~\textbf{\textit{-6.5153}} \\
~\multirow{2}{*}{\underline{-0.9485}}   & ~-1.8365 & ~\textit{-2.8667} &  & ~\multirow{2}{*}{\underline{0.3568}}  & ~\textit{-0.3866} & ~-1.2587 & & ~\multirow{2}{*}{\underline{2.9565}} & ~2.4852 & ~\textit{1.4012} & ~\textbf{1.3968} \\
 & ~0.5636 & ~0.9116 & ~\multirow{2}{*}{\underline{2.3243}} &  & ~\textit{1.5817} & ~\textit{3.0960}& ~\multirow{2}{*}{\underline{4.2156}} &  & ~3.6567 & ~6.7965 & ~\multirow{2}{*}{\underline{7.4914}}\\
~\multirow{2}{*}{\underline{5.0669}} & ~\textit{2.6724} & ~4.8582 &  & ~\multirow{2}{*}{\underline{6.1324}} & ~4.0040 & ~\textit{6.2819}&  & ~\multirow{2}{*}{\underline{8.3940}} & ~\textit{6.7831} & ~8.7353 &  \\
  & ~\textit{8.7349} & ~\textit{5.9178} & ~\multirow{2}{*}{\underline{9.2611}} &   & ~9.5751 & ~7.6661 & \multirow{2}{*}{\underline{10.6232}} &  & \textit{11.3059} & \textit{10.8284} & \multirow{2}{*}{\underline{13.0684}} \\
 \multirow{2}{*}{\underline{13.4317}}& 9.1076 & \textit{13.6472} & & \multirow{2}{*}{\underline{14.1848}} & \textit{10.0361} & 14.7668 &  & \multirow{2}{*}{\underline{15.7486}} & 12.0093 &\textit{16.7074}&  \\
& 18.4942 & 13.7747 & \multirow{2}{*}{\underline{18.7515}} &  & \textit{19.1368} & \textit{14.9465} & \multirow{2}{*}{\underline{19.7328}} & & 20.4547 & 17.0293 & \multirow{2}{*}{\underline{21.4336}}\\[2ex]
\end{tabular}
\end{ruledtabular}
\caption{Analytical (bold) and numerical eigenenergies, $E_{t,\kappa}^{(\Gamma_t)}$, of the pendular (trigonometric) Hamiltonian $H_t$, for $\beta=-5$. Note that energy values of odd ($A_2$ and $B_2$ representations) states are shown in italics. Underlined values indicate degenerate states (doublets of even and odd states). Numerical values obtained by \textsc{WavePacket} software \cite{BSchmidt:75}.}
\label{tab:num_pendulum}
\end{table} 
\end{turnpage}

%RRRRRRRRRRRRRRRRRRRRRRRRRRRRRRRRRRRRRRRRRRRRRRRRRRRRRRRR
\begin{table}
\begin{ruledtabular}
\begin{tabular}{cccccc}
$\kappa=1$ & $\kappa=2$ & $\kappa=3$ & $\kappa=4$ & $\kappa=5$ & $\kappa=6$\\
\hline
\textbf{25}      & \textbf{19.75}            & \textbf{14.4875}          & \textbf{9.2106}           &\textbf{3.9169}           &\textbf{-1.3968}          \\
\textit{35.4684} & \textbf{\textit{29.75}}   & \textbf{\textit{24}}      & \textbf{\textit{18.2143}} &\textbf{\textit{12.3881}} &\textbf{\textit{6.5153}}  \\
46.8234          & 40.6891                   & \textbf{34.5125}          & \textbf{28.2894}          &\textbf{22.0150}          &\textbf{15.6840}          \\ 
\textit{58.9796} & \textit{52.4654}          & \textit{45.9020}          & \textbf{\textit{39.2857}} &\textbf{\textit{32.6119}} &\textbf{\textit{25.8760}} \\
71.8748          & 65.0075                   & 58.0864                   &  51.1079                  &\textbf{44.0681}          &\textbf{36.9628}          \\ 
\textit{85.4614} & \textit{78.2621}          & \textit{71.0055}          & \textit{63.6885}          &\textit{56.3074}          &\textbf{\textit{48.8587}} \\
99.7011          & 92.1867                   & 84.6127                   & 76.9758                   & 69.2730                  & 61.5010                  \\
\textit{114.5623}& \textit{106.7472}         & \textit{98.8704}          & \textit{90.9291}          &\textit{82.9204}          &\textit{74.8415}          \\
130.0184         & 121.9146                  & 113.7476                  & 105.5147                  & 97.2135                  &  88.8411                 \\
\textit{146.0465}& \textit{137.6644}         & \textit{129.2180}         & \textit{120.7047}         &\textit{112.1221}         &\textit{103.4679}         \\
162.6266         & 153.9756                  & 145.2591                  & 136.4749                  & 127.6209                 & 118.6947                 \\ 
\end{tabular}
\end{ruledtabular}
\caption{Analytical (bold) and numerical eigenenergies, $E_{h,\kappa}^{(\Gamma_h)}$, of the Razavy (hyperbolic) Hamiltonian $H_h$ with $\beta=-5$. Note that energy values of odd ($A''$ representation) states are shown in italics. The numerical values were obtained with \textsc{WavePacket} software \cite{BSchmidt:75}.}
\label{tab:num_razavy}
\end{table}

\clearpage

\begin{figure}
\includegraphics[width=8cm]{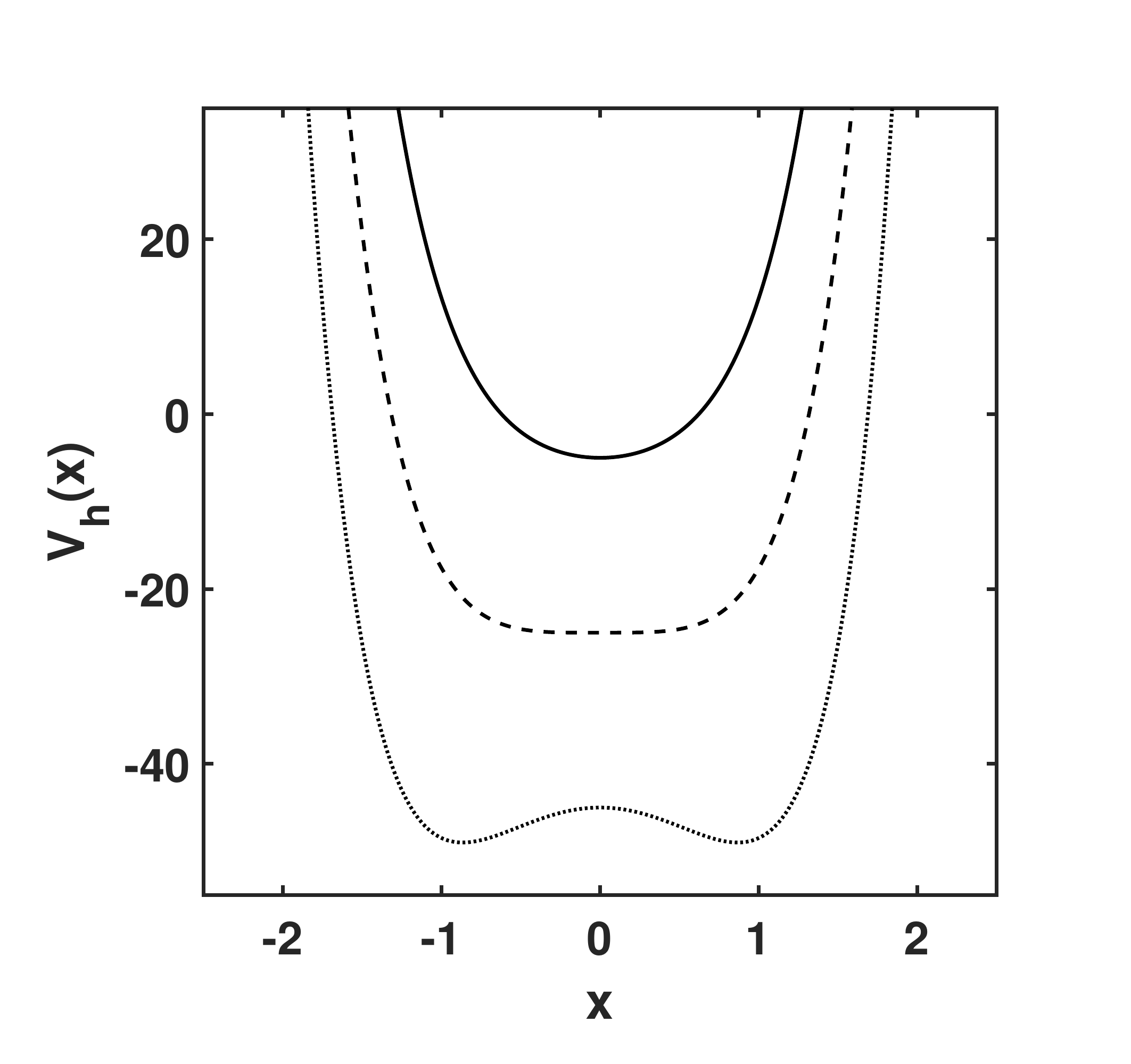}
\caption{Razavy (hyperbolic) potential (\ref{eq:pot_h}) for $\zeta=25$ and $\eta=-70$ (dotted curve), $\eta=-50$ (dashed), and $\eta=-30$ (full). The potential is a double well for $\left|\eta\right| \geq \left|2\zeta\right|$ and a single well for $\left|\eta\right| < \left|2\zeta\right|$}
\label{fig:razavy}
\end{figure} 

\begin{figure}
\includegraphics[width=17cm]{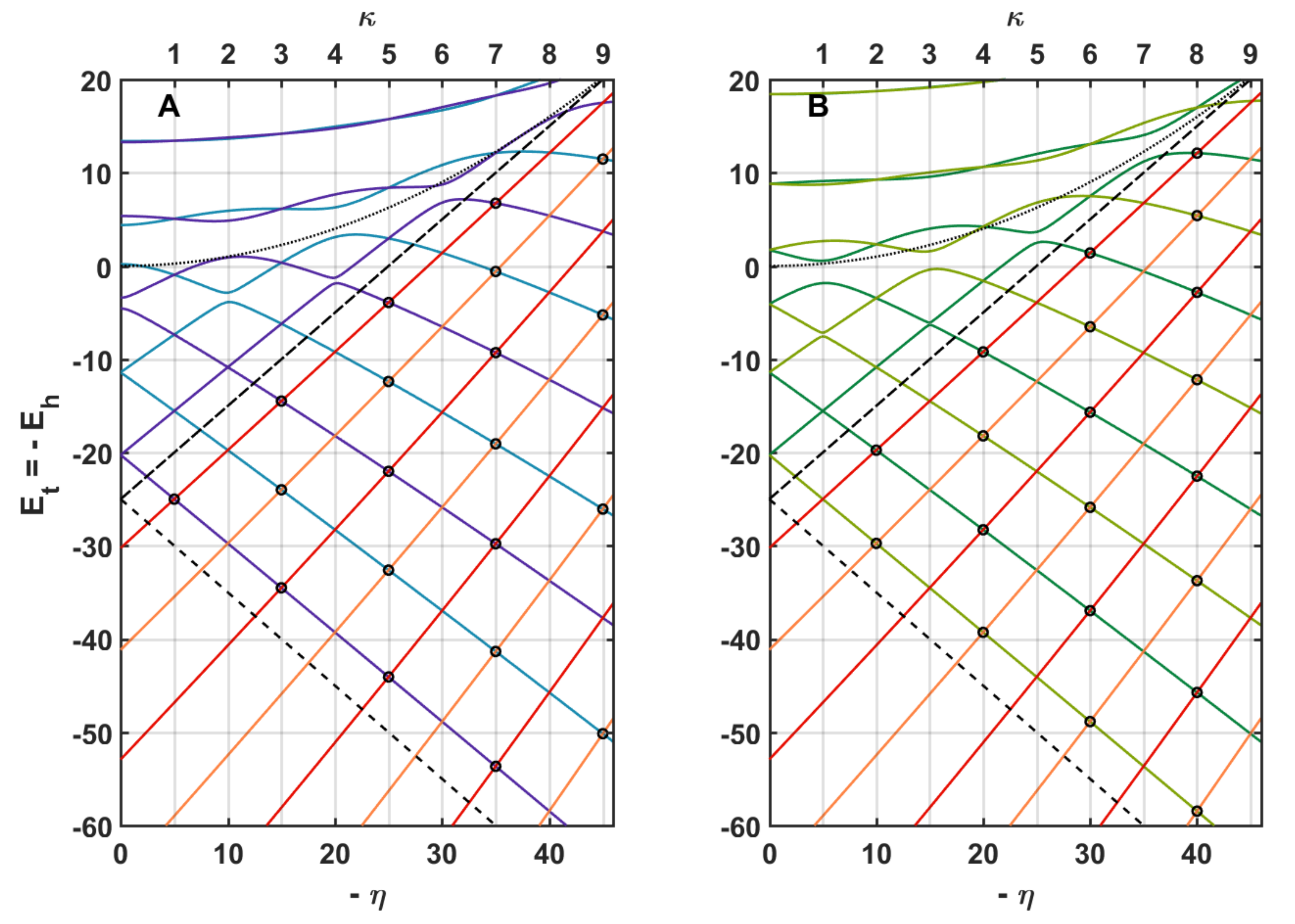}
\caption{Periodic (panel A) and anti-periodic (panel B) energies of the planar pendulum and inverted energies of the Razavy system for $\beta=-5$. For this choice of $\beta$, $V_t$ is a double well with a local minimum (thick dashed line) of $V_t(\theta_{min,l})$, a global minimum (dashed line) and a maximum (dotted  curve); $V_h$ is a single well potential whose minimum (shown by the thick dashed line) is $V_t(\theta_{min,l}) = - V_h(x_{min})$.  The colors follow the scheme introduced in Fig.~\ref{fig:seed}. Circles show analytic eigenenergies, which coincide for the trigonometric and hyperbolic systems. The numerical values were obtained with \textsc{WavePacket} software \cite{BSchmidt:75}.}
\label{fig:e_b500}
\end{figure}

\begin{figure}
\includegraphics[width=17cm]{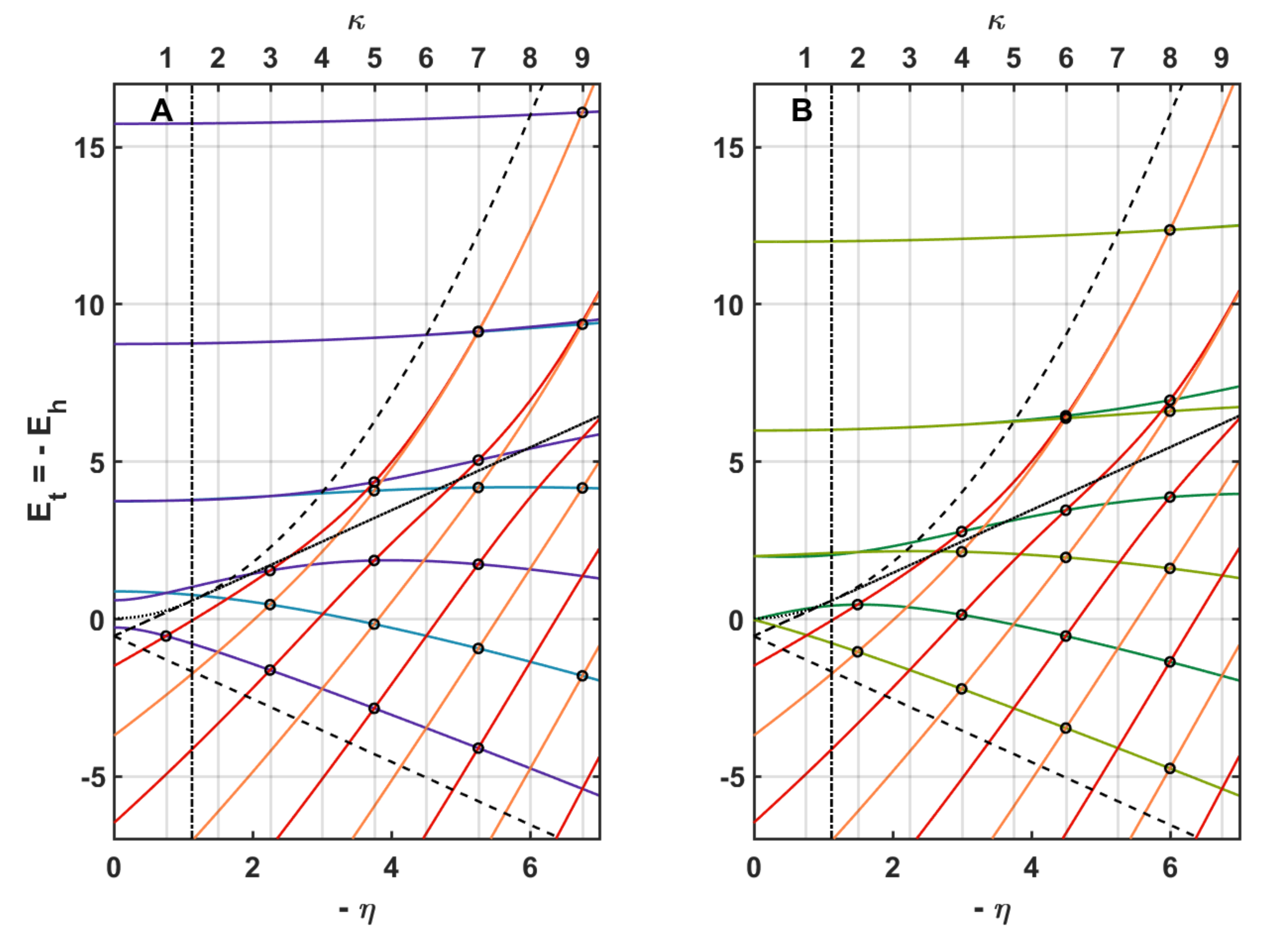}
\caption{Periodic (panel A) and anti-periodic (panel B) energies of the planar pendulum and inverted energies of the Razavy syatem, for $\beta=-3/4$. On the left side of the vertical (dash-dotted) line, the potentials are qualitatively the same as in Fig.~\ref{fig:e_b500}. On the right side, $V_t$ is a single well potential with a minimum (dashed line) and a maximum (thick dotted line) whereas $V_h$ is a double well potential with two equal minima (dashed curve) and one maximum (thick dotted line). The colors follow the scheme introduced in Fig.~\ref{fig:seed}. Circles show analytic eigenenergies, which coincide for the trigonometric and hyperbolic systems. The numerical values were obtained with \textsc{WavePacket} software \cite{BSchmidt:75}.} 
\label{fig:e_b075}
\end{figure}

\end{document}